\documentclass[times,twocolumn,final]{elsarticle}

\usepackage{cag}
\usepackage{framed,multirow}

\usepackage{amssymb}
\usepackage{latexsym}

\newtheorem{definition}{Definition}
\usepackage{amsmath}
\usepackage{subfigure}
\newcommand{\reffig}[1]{Fig.\ref{#1}}
\usepackage{float}
\usepackage{titlesec}
\usepackage[titletoc]{appendix}
\usepackage[noend]{algpseudocode}
\usepackage{algorithmicx,algorithm}

\usepackage{url}
\usepackage{xcolor}
\definecolor{newcolor}{rgb}{.8,.349,.1}

\usepackage{hyperref}

\usepackage[switch,pagewise]{lineno} 

\journal{Arxiv}

\begin{document}

\verso{Preprint Submitted for review}

\begin{frontmatter}

\title{Golden interpolation\tnoteref{tnote1}}%
\tnotetext[tnote1]{Only capitalize first
word and proper nouns in the title.}

\author[1]{Ying \snm{He}}
\author[2]{Jincai \snm{Chang}\corref{cor1}}
\cortext[cor1]{Corresponding author:
  Tel.: +8615100511216;}
\ead{jincai@ncst.edu.cn}

\address[1]{College of Information Engineering, North China University of Science and Technology, Tangshan 063210, China}
\address[2]{College of Sciences, North China University of Science and Technology, Tangshan 063210, China}

\received{\today}

\begin{abstract}
For the classic aesthetic interpolation problem, we propose an entirely new thought: apply the golden section. For how to apply the golden section to interpolation methods, we present three examples: the golden step interpolation, the golden piecewise linear interpolation and the golden curve interpolation, which respectively deal with the applications of golden section in the interpolation of degree 0, 1, and 2 in the plane. In each example, we present our basic ideas, the specific methods, comparative examples and applications, and relevant criteria. And it is worth mentioning that for aesthetics, we propose two novel concepts: the golden cuspidal hill and the golden domed hill. This paper aims to provide the reference for the combination of golden section and interpolation, and stimulate more and better related researches.
\end{abstract}


\end{frontmatter}


\section{Introduction}

The golden section has a long history: the earliest discovery can be traced back to the Greek Pythagoras School in the sixth century BC, and then be systematically studied by the Greek mathematician Eudoxus in the fourth century BC, and be given a complete geometry definition and proof in the Euclidean Geometry in the third century BC~\cite{r1gold}. The golden section is mysterious and is found related to many aesthetic things in the nature~\cite{r2gold}, such as our body~\cite{re5gold}, the canopy structure of forage plants~\cite{re6gold} and so on. In mathematics, people are also surprised to find the widespread existence of the golden section. For example, the literature~\cite{re7gold} describes the discovery that super central configurations of the n-body problem have surprising connections with the golden section ratio. For another example, the literature~\cite{r7splinegold} points out that the golden section exists in the spline function. Since its birth, golden section has been connected to beauty and has become a universally recognized aesthetic law~\cite{r3goldbeauty}. Researches show that the aesthetic experience of golden section is not entirely subjective and there are an objective biological basis~\cite{re4gold} and some psychological basis~\cite{re8gold}. Though there has been no definite explanation for why the golden section is beautiful, it is found that it has repeatedly played an effective role in practice. Nowadays, the aesthetic value of golden section has been widely applied in the fields of art, nature, architecture, mathematics, philosophy, production practice and so on. The golden section ratio is consistently represented by the Greek letter $\varphi$, and so is it in the following sections of this paper.

Interpolation is a classic problem. In the interpolation problem, it is required to construct a function that exactly matches prescribed data, such as a series of points and derivatives. Common interpolation methods include the nearest neighbor interpolation, the spline interpolation, the Lagrange interpolation, the Hermite interpolation and so on~\cite{r4interpolation,r5interpolation,r6spline}. Nowadays interpolation methods are used not only for estimation, but also for computer-aided design, shape modeling, geometric processing and many other fields~\cite{r8cagd} in which the appearance of the interpolation function graph is of great importance. As a result, aesthetic interpolation forms a fundamental problem to explore a better appearance. There have been lots of researches struggling for it. In general, the existing methods are in accordance with the following thoughts: (1) Improve the fairness of the interpolation curve(surface). This thought was in embryo when it was 1966~\cite{energyembryo}, and then it progressively developed~\cite{energydevelop1,energydevelop2}. In 1994, Nowacki and L{\"u}~\cite{energycriterion} went further and adopted a fairness criterion about it. Nowadays, minimizing some energy functions becomes a universally accepted approach to improve the fairness of the interpolation, in which the strain energy(also called bending energy) and curvature variation energy are mostly used, e.g. the literatures~\cite{rfair2strainEnergy,rfair3strainEnergy,rfair1curvature,rfair2curvature}. (2) Interpolate with a log-aesthetic interpolation curve. This thought can date back to 1997 when Harada analyzed many aesthetic curves including feature curves of automobiles and then proposed the log-aesthetic curves~\cite{logfairfind}. Nowadays this thought has also been widely applied to enhance the beauty of interpolation or approximation function graphs, e.g. the literatures~\cite{logfair1,logfair2,logfair3}. (3) If a curve has monotonous curvature, then it is deemed more pleasant. Spirals have monotonous curvature, and interpolating with a spiral gradually becomes a criterion for aesthetic interpolation. So there is lots of work studying the aesthetic interpolation according to this thought, e.g. the literatures~\cite{spiral1,spiral2,spiral3}.

As a famous aesthetic law, golden section has been widely used to enhance the beauty of things, but as far as we know, no one has applied the golden section principle to solve the aesthetic interpolation problem. The literature~\cite{goldenCAD} studied the approximation of spirals by piecewise curves of fewest circular arc segments, in which there existed an application of the golden section, but it was used for the extremal value finding in iterative computation rather than for changing the appearance of the curve. In this paper, we propose a novel thought for the fundamental aesthetic interpolation problem: apply the golden section principle to improve the beauty of the interpolation function graph. We call this kind of interpolation ``golden interpolation''. In the following of this paper, when we use the adjective ``golden'' before a concept, we mean there is the application of golden section to it. However, how do we apply golden section to interpolation methods? For this question, we start from the simplest case, and then gradually go more complex, presenting three examples: the golden step interpolation, the golden piecewise linear interpolation and the golden curve interpolation, which respectively discuss how to apply golden section to three common interpolation methods: the one-dimensional interpolation of degree 0, 1, and 2. In each example, we present our basic ideas, elaborate on specific methods, demonstrate comparative numerical examples and applications, and explore relevant criteria. Golden interpolation is meaningful for modeling, design, graphics and other fields. This paper aims to attract attention to the application of golden section in interpolation, provide related references, and stimulate better methods.

The rest of this paper is organized as follows. In Section~\ref{sec:transform}, we propose controlling interpolation by adding a node transform, which serves our golden methods. In Section~\ref{sec:interpolation}, we present the traditional methods that are used in this paper. In Section~\ref{sec:step}, we propose the golden step interpolation. In Section~\ref{sec:linear}, we we propose the golden piecewise linear interpolation. In Section~\ref{sec:curve}, we propose the golden curve interpolation. And finally, in Section~\ref{sec:conclusion}, we conclude this paper and propose some prospects. For the convenience of the computer implementation, we describe our methods in the form of algorithms by using pseudo codes and put them in Appendix~\ref{appendix:algorithm}.


\section{Control interpolation by adding a node transform}\label{sec:transform}

Since the geometric characteristics of many interpolation function graphs are clearly related to the interpolation data nodes, we can easily change the function graph into the shape we want by transforming the data nodes. So in this paper, we do interpolation following the two steps: first, transform the data nodes \({A_0},{A_1}, \cdots,{A_n}\) into \({B_0},{B_1}, \cdots,{B_m}\); second, construct a traditional interpolation function for \({B_0},{B_1}, \cdots,{B_m}\).

On considering the demand for this paper, we define two kinds of data node transforms:

\begin{definition}\label{def:extrsfm}
 If the number of data nodes \({A_0},{A_1}, \cdots,{A_n}\) is smaller than the number of data nodes \({B_0},{B_1}, \cdots,{B_m}\), then the transform is called an extension transform.
\end{definition}

\begin{definition}\label{def:eqtrsfm}
If the number of data nodes \({A_0},{A_1}, \cdots,{A_n}\) is equal to the number of data nodes \({B_0},{B_1}, \cdots,{B_m}\), then the transform is called an equal number transform.
\end{definition}

\section{Spline interpolation of degree 0, 1, and 2}\label{sec:interpolation}
We construct a golden interpolation function by adding a transform to the traditional method. In this paper, the following traditional methods have been used.

Give interpolation data nodes \(A_0(x_0,y_0)\), \(A_1(x_1,y_1)\), \(\cdots\) , \(A_n(x_n,y_n)\), where \({x_0} < {x_1} < \cdots <{x_n}\), \(n\ge 1\). And for quadratic spline interpolation, we also need to give the derivative \({{k}_{0}}\) at the node \({{A}_{0}}\).

\begin{enumerate}[(1)]

\item Zero-degree spline interpolation

The interpolation function is \[p(x) = {y_i},\] where \(x \in [{x_i},{x_{i + 1}})\), \(i = 0,1, \cdots,n - 1\).

\item Piecewise linear interpolation

The interpolation function is \[p(x) = \frac{x - x_{i + 1}}{x_i - x_{i + 1}}y_i + \frac{x - x_i}{{x_{i + 1} - x_i}}y_{i + 1}, \] where \(x \in [{x_i},{x_{i + 1}}],i = 0,1, \cdots,n - 1\).

It is also the one-degree spline interpolation.

\item Quadratic spline interpolation

The interpolation function is
\[\begin{split}
p(x) = &{\left(\frac{x - x_{i + 1}}{x_i - x_{i + 1}}\right)}^{2}y_i + {\left(\frac{x - x_i}{{x_{i + 1} - x_i}}\right)}^{2}y_{i + 1} + \\
&\frac{(x - x_i)(x - x_{i + 1})}{x_i - x_{i + 1}}\Bigg[2\sum_{j=0}^{i-1}(-1)^{i-j+1}\frac{y_{j + 1} - y_j}{x_{j + 1} - x_j} + \\
&{(-1)^i}{k_0} - \frac{2y_i}{x_i - x_{i + 1}}\Bigg],
\end{split} \]
where \(x \in [{x_i},{x_{i + 1}}],i = 0,1, \cdots,n - 1\).

It is also the two-degree spline interpolation.
\end{enumerate}

A quadratic spline is composed of piecewise quadratic polynomials with continuous first-order derivatives at spline knots. In order to write its interpolation function according to interpolation data nodes directly, we create the above formula referring to the thought of barycentric rational Hermite interpolation~\cite{Kai2018Barycentric}. Here we elaborate how we educe the formula above.

Consider the interval \([x_i,x_{i+1}]\). We all know the two basis functions:

\centerline{\(b_i(x)=\displaystyle\frac{x - x_{i + 1}}{x_i - x_{i + 1}} \) , \(b_{i+1}(x)=\displaystyle\frac{x - x_i}{{x_{i + 1} - x_i}}\).}

They have the properties:

\centerline{\(b_i(x)=\begin{cases}1, x=x_i,\\0, x=x_{i+1},\end{cases}\)  \(b_{i+1}(x)=\begin{cases}0, x=x_i,\\1, x=x_{i+1}.\end{cases}\)}

It is well-known that the linear combination \(b_i(x){y_i} + b_{i+1}(x){y_{i+1}}\) is a one-degree polynomial that passes through \(A_i\) and \(A_{i+1}\). Then we can easily think out a two-degree polynomial that passes through \(A_i\) and \(A_{i+1}\):\[r_i(x)=b_i(x)^2{y_i} + b_{i+1}(x)^2{y_{i+1}}.\]

But for the quadratic spline, we also need to restrict the derivative at \(x_i\), so that the first-order derivative of the whole function at \(x_i\) is continuous. Referring to the Hermite interpolation~\cite{Kai2018Barycentric}, we can think out the following expression:
\begin{equation}
p_i(x)=r_i(x) + (x - x_i)b_i(x)[p'_{i-1}(x_i) - r'_i(x_i)].
\label{con:qspline}
\end{equation}

It's easy to work out: \[p_i(x_i)=y_i,\] \[p_i(x_{i+1})=y_{i+1},\] \[p'_i(x_i)=p'_{i-1}(x_i),\] \[p'_i(x_{i+1})=\frac{2(y_{i+1} - y_i)}{x_{i + 1} - x_i} - p'_{i-1}(x_i).\]

We mark \(q_i=\displaystyle\frac{y_{i+1} - y_i}{x_{i + 1} - x_i}\). Then \(p'_i(x_{i+1})=2q_i - p'_{i-1}(x_i)\),
\[\begin{split}
p'_{i-1}(x_i) &= 2q_{i-1} - p'_{i-2}(x_{i - 1})\\
&= 2q_{i-1} - [2k_{i-2} - p'_{i-3}(x_{i - 2})]\\
&= 2q_{i-1} - 2q_{i-2} + 2q_{i-3} - \cdots +(-)2q_0 -(+) k_0\\
&= 2\sum_{j=0}^{i-1}(-1)^{i-j+1}q_j + (-1)^i{k_0}.
\end{split} \]

Put all the equations into the equation~\eqref{con:qspline} and we can get the expression of the quadratic spline interpolation function above.

\section{Golden step interpolation}\label{sec:step}

Among one-dimensional interpolation methods, the step function interpolation is the simplest and plainest. So this paper first discusses how to apply the golden section principle to the step function interpolation. The graph of the step interpolation function is a set of horizontal line segments that respectively pass through the given data nodes, and the lengths of the line segments are the most obvious geometric feature of the graph. So the basic idea of our golden step interpolation is trying to make the ratio of the lengths of the two adjacent segments satisfy the golden section ratio. Common the step function interpolation methods are various, like the nearest neighbor interpolation, the zero-degree spline interpolation, etc. Considering the zero-degree spline interpolation has the property that we can easily change the function graph into the shape we want by transforming the interpolation data nodes, we use this traditional method as a basis, and add a golden transform for the data nodes to enhance the beauty of the final interpolation graph.

\subsection{Golden extension step interpolation}\label{sec:stepGE}
We add an extension transform to the zero-degree spline interpolation, in which additional nodes are added to the original interpolation node set, and finally makes the ratio of the lengths of two line segments between two adjacent original interpolation nodes satisfy the golden section ratio. The specific interpolation process includes two steps as follows, and we call this method the golden extension step interpolation.

\begin{enumerate}[step 1.]

\item Transform \({A_0}\), \({A_1}\), \(\cdots\), \({A_n}\) into \({{A}_{0}}\), \({{A}_{0.5}}\), \({{A}_{1}}\), \({{A}_{1.5}}\), \(\cdots\) , \({{A}_{n-0.5}}\), \({{A}_{n}}\) as follows:

Add a data node \({{A}_{i+0.5}}({{x}_{i+0.5}},{{y}_{i+0.5}})\) between every two adjacent nodes \({{A}_{i}},{{A}_{i+1}}(i=0,1,\cdots,n-1)\), in which \[{{x}_{i+0.5}}={{x}_{i+1}}-({{x}_{i+1}}-{{x}_{i}})\varphi, \] and \[{y_{i + 0.5}} = \left\{
\begin{aligned}
&{y_{i + 1}} - ({y_{i + 1}} - {y_i})\varphi ,{y_i} \ne {y_{i + 1}},\\
&{y_{i + 1}} + L,{y_i} = {y_{i + 1}}.
\end{aligned}
\right.\]\(L(L \neq 0)\) is a longitudinal jump that is given \({{A}_{i+0.5}}\) when \({{y}_{i}}={{y}_{i+1}}\) and it should be determined according to the actual situation.

\item Get the final interpolation function \({{p}_{se}}(x)={{y}_{0.5i}}\), where \(x\in [{{x}_{0.5i}},{{x}_{0.5(i+1)}}),i=0,1,\cdots,2n-1\).

\end{enumerate}

We need to explain that there are two golden section points in an interval, but here we take the left one. We can also take the right point by replacing the \(\varphi\) with \(1 - \varphi\). So is it for other methods in this paper and we won't explain it again.

\subsection{Golden equal number step interpolation}\label{sec:stepG}
The above golden extension step interpolation will extend data nodes. Although it can guarantee that the ratio of the lengths of two line segments between two adjacent interpolation data nodes satisfies the golden section ratio in the graph, compared with the direct zero-degree spline interpolation, this method increases the discontinuities of the function, and the number of line segments in the graph is doubled, and the length of each line segment is shortened. However, sometimes we want to make the function graph look more beautiful on condition that the interpolation function has the same number of discontinuities as the direct zero-degree spline interpolation function. Therefore, we also propose adding a golden equal number transform, which makes the ratio of the lengths of two adjacent line segments in the final interpolation function graph satisfy the golden section ratio as far as possible. The specific interpolation process includes two steps as follows, and we call this method the golden equal number step interpolation.

\begin{enumerate}[step 1.]

\item Transform \({{A}_{0}},{{A}_{1}},\cdots,{{A}_{n}}\) into \({{A}_{0}},{{{A}'}_{1}},{{A}_{2}},{{{A}'}_{3}},\cdots,{{A}_{n}}\) as follows:

If \(n<2\), the data nodes are not changed; that is, the transformed data nodes are the same as the original data nodes.

If \(n\ge 2\), take out the data nodes \({{A}_{2i-1}}({{x}_{2i-1}},{{y}_{2i-1}}),i=1,2,\cdots,[0.5n]\) whose sequence numbers are odd numbers among the \(n\) data nodes, in which \([0.5n]\) represents the integer part of \(0.5n\). Then transform the x-coordinate \({{x}_{2i-1}}\) of \({{A}_{2i-1}}({{x}_{2i-1}},{{y}_{2i-1}})\) into \[{x'_{2i - 1}} = \left\{
\begin{aligned}
&{x_g},{x_g} < {x_{2i - 1}},\\
&{x_{2i - 1}},{x_g} \ge {x_{2i - 1}},
\end{aligned}
\right. \]where \({x_g}={{x}_{2i}}-({{x}_{2i}}-{{x}_{2i-2}})\varphi \).

Then the data node \({{A}_{2i-1}}({{x}_{2i-1}},{{y}_{2i-1}})\) is transformed into \({{{A}'}_{2i-1}}({{{x}'}_{2i-1}},{{y}_{2i-1}})\). The meaning of this transform will be explained below.

\item For convenience, we mark \({{A}_{0}},{{{A}'}_{1}},{{A}_{2}},{{{A}'}_{3}},\cdots,{{A}_{n}}\) as \(A'_0(x'_0,y_0)\), \(A'_1(x'_1,y_1)\), \(A'_2(x'_2,y_2)\), \(A'_3(x'_3,y_3)\), \(\cdots\) , \(A'_n(x'_n,y_n)\), and get the final interpolation function \(p_s(x)=y_i\), where \(x\in [x'_i,x'_{i+1}),i=0,1,\cdots,n-1\).

\end{enumerate}

It is necessary to explain the meaning of the transform in step 1. The meaning is that if we can make the ratio of the length of the line segment passing through \({{A}_{2i-2}}\) to the length of the line segment passing through \({{A}_{2i-1}}\)  satisfy the golden section ratio in the function graph, then change the data node \({{A}_{2i-1}}\), otherwise keep the data node \({{A}_{2i-1}}\) unchanged. So this transform just try to be as far as possible and the transformed data nodes are not always different from the original data nodes.

\subsection{Examples and applications}
Give data nodes \(B_0(2,1)\), \(B_1(9,3)\), \(B_2(11,4)\), \(B_3(19,6)\), \(B_4(21,2)\), \(B_5(24,5)\).

\begin{figure*}[htbp] \centering
\subfigure[] { \label{fig:stair}
\includegraphics[width=0.8\columnwidth]{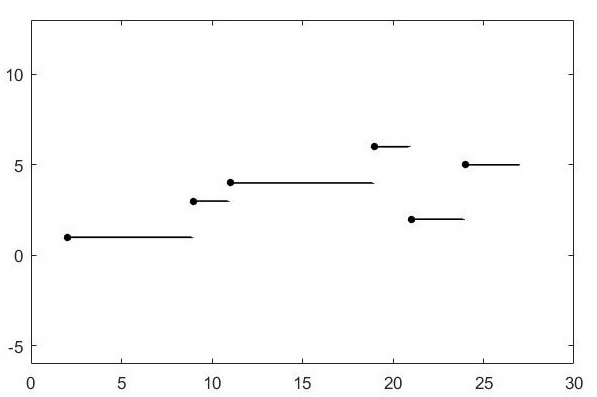}
}
\subfigure[] { \label{fig:stairG}
\includegraphics[width=0.8\columnwidth]{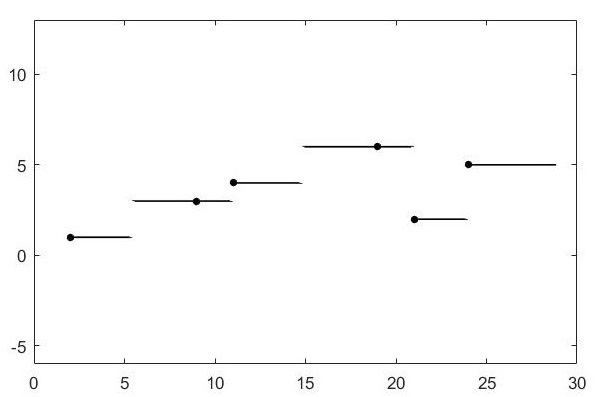}
}
\\
\subfigure[] { \label{fig:stairGApp}
\includegraphics[width=0.75\columnwidth]{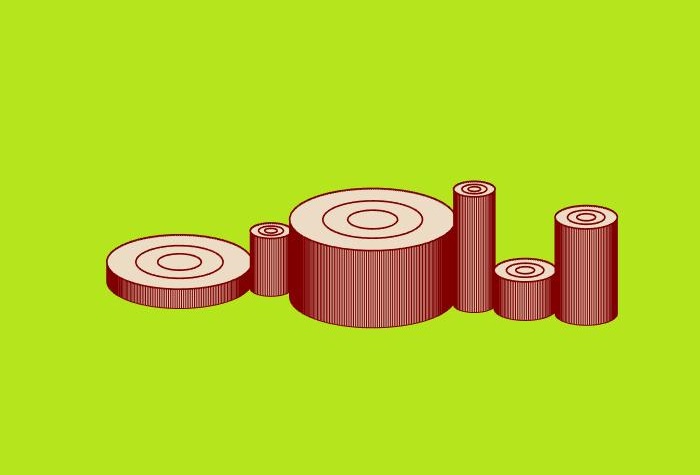}
}
\hspace{11pt}
\subfigure[] { \label{fig:stairApp}
\includegraphics[width=0.75\columnwidth]{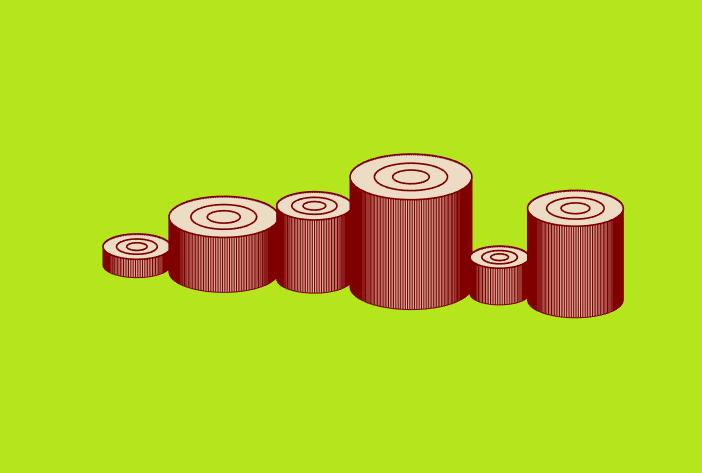}
}
\caption{ Comparative examples and applications: (a) the graph of the traditional zero-degree spline interpolation function; (b) the graph of the golden equal number step interpolation function; (c) the 3D stumps generated by (a); (d) the 3D stumps generated by (b). }
\label{fig:stairGall}
\end{figure*}

Do the traditional zero-degree spline interpolation and the golden equal number step interpolation for them. The function graphs are~\reffig{fig:stair} and~\reffig{fig:stairG}. In order to exist a line segment passing through the final interpolation data node, we extend the definition domain of the function, such that for the line segment passing through \(B_5\), its length equals the length of its previous adjacent line segment in~\reffig{fig:stair} and is \(1/\varphi\) times the length of its previous adjacent line segment in~\reffig{fig:stairG}. Then we discuss an application of designing the stumps in the park: If we want to design some cylindrical stumps next to each other with diverse diameters and heights in the park for decorating, exercising and playing, the designer only needs to point some nodes on the panel to decide the approximate locations and heights of the stumps. We also suppose these nodes are the \(B_0,B_1,\cdots,B_5\) above. Then the computer can automatically generate the line segments in~\reffig{fig:stair} or~\reffig{fig:stairG}. The side view of these stumps can be constructed by adding the perpendicular lines of the horizontal axis through the ends of these line segments. Rotate each side view of these stumps and the 3D model of the stumps can be built. See~\reffig{fig:stairApp} and~\reffig{fig:stairGApp}. By comparing, we can observe that the appearance generated by the golden method is more harmonious and aesthetic than the one generated by the traditional method. It is owed to the integration of gold ratio in geometric features in the golden method.

\begin{figure*}[htbp] \centering
\subfigure[] { \label{fig:stairGE}
\includegraphics[width=0.79\columnwidth]{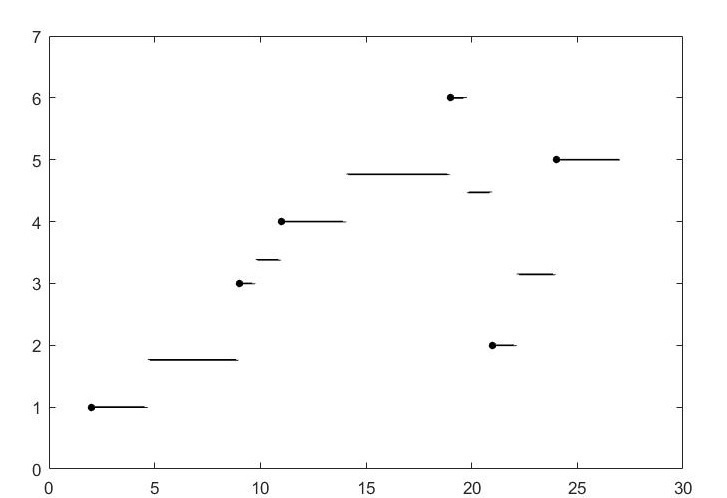}
}
\subfigure[] { \label{fig:stairGEApp}
\includegraphics[width=0.75\columnwidth]{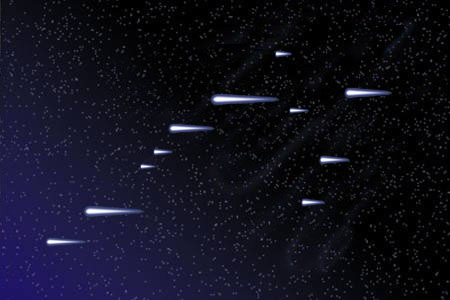}
}
\caption{ An example and an application of the golden extension step interpolation: (a) the function graph; (b) an application of designing a picture in which meteorites are streaking across the night sky. }
\label{fig:gstairGEall}
\end{figure*}

For the golden extension step interpolation, the function graph interpolating the above nodes is~\reffig{fig:stairGE}, in which the length of the final line segment is \(1/\varphi\) times that of the previous one. For this golden method, we can use it to design a picture in which meteorites are streaking across the night sky: The designer only needs to point several nodes on the canvas. Let's suppose these nodes are the \(B_0,B_1,\cdots,B_5\) above. Then the computer can automatically generate the tracks of the meteorites passing through these nodes in the night sky. These tracks are the line segments in~\reffig{fig:stairGE}, and the generated picture is shown in~\reffig{fig:stairGEApp}. We can see the picture~\reffig{fig:stairGEApp} looks harmonious and aesthetic.

\subsection{Go further to explore criteria}\label{subsec:criteriaStep}
For the aesthetic interpolation problem, most of the traditional thoughts have developed some criteria. For example, strain energy and curvature variation energy have been universally used to measure the fairness of a curve~\cite{rfair2strainEnergy,rfair3strainEnergy,rfair1curvature,rfair2curvature}, which can provide reference for new interpolation methods and turn the aesthetic interpolation problem into an optimization problem to solve. For our golden interpolation, it is also expected to explore some criteria to measure the golden degree of the function graph, and to analyse the golden methods mathematically and quantificationally. So we do the following exploration for the golden step interpolation.

Suppose the \(n+1\) nodes \((x_0,y_0)\), \((x_1,y_1)\), \(\cdots\) , \((x_n,y_n)\) are the nodes directly input to the zero-degree spline interpolation and any two of \(y_0, y_1, \cdots, y_n\) are unequal. Consider the \(m+1(m \ge 2)\) adjacent nodes \((x_d,y_d)\), \((x_{d+1},y_{d+1})\), \(\cdots\) , \((x_{d+m},y_{d+m})\), where \(d = 0, m, 2m, \cdots, lm\), and \(l = [n/m] - 1\), where the ``[]" represents taking the integer part. Then there are \(m\) line segments in the function graph, whose lengths are \(x_{d+1} - x_d\), \(x_{d+2} - x_{d+1}\), \(\cdots\) , \(x_{d+m} - x_{d+m-1}\) respectively. The following criteria are easy to explore.

\begin{enumerate}[(1)]

\item Left golden error

If every \(m\) line segments constitute a group and we hope the ratio of the lengths of any two adjacent line segments satisfies \(\varphi\) in each group, then the following \(E_{left}\) can measure the error between present graph and the goal.

\[\begin{split}
E_{left} = &\sum_{i=0}^{l}\sum_{j=0}^{m-2}\left| \frac{x_{im+j+1} - x_{im+j}}{x_{im+j+2} - x_{im+j}} - (1 - \varphi) \right| + \\
&\sum_{j=m(l+1)}^{n-2}\left| \frac{x_{j+1} - x_j}{x_{j+2} - x_j} - (1 - \varphi) \right|
.\end{split}\]

In this formula, when the upper limit is smaller than the lower limit of \(\sum\), the sum is 0. And so is it in the rest of this paper.

\item Right golden error

If we hope the ratio is \(1 / \varphi\), then the following \(E_{right}\) can measure the error.

\[\begin{split}
E_{right} = &\sum_{i=0}^{l}\sum_{j=0}^{m-2}\left| \frac{x_{im+j+1} - x_{im+j}}{x_{im+j+2} - x_{im+j}} - \varphi \right| + \\
&\sum_{j=m(l+1)}^{n-2}\left| \frac{x_{j+1} - x_j}{x_{j+2} - x_j} - \varphi \right|
.\end{split}\]

\item Mixed golden error

If the ratio is smaller than 1, we hope it is \(\varphi\); Otherwise, we hope it is \(1 / \varphi\). Then the following \(E_{mixed}\) can measure the error.

\[\begin{split}
E_{mixed} = &\sum_{i=0}^{l}\sum_{j=0}^{m-2}\left| \frac{x_{im+j+1} - x_{im+j}}{x_{im+j+2} - x_{im+j}} - q_{i,j} \right| + \\
&\sum_{j=m(l+1)}^{n-2}\left| \frac{x_{j+1} - x_j}{x_{j+2} - x_j} - q_j \right|
.\end{split}\]

where
\[q_{i,j} =\left\{
\begin{aligned}
&1 - \varphi, {x_{im+j+1} - x_{im+j}} < {x_{im+j+2} - x_{im+j+1}},\\
&\varphi, {x_{im+j+1} - x_{im+j}} \ge {x_{im+j+2} - x_{im+j+1}},
\end{aligned}
\right.\]
and
\[q_j = \left\{
\begin{aligned}
&1 - \varphi, {x_{j+1} - x_j} < {x_{j+2} - x_{j+1}},\\
&\varphi, {x_{j+1} - x_j} \ge {x_{j+2} - x_{j+1}}.
\end{aligned}
\right.\]

\item Alternate golden error

If we hope the ratios are \(\varphi\), \(1 / \varphi\), \(\varphi\), \(1 / \varphi\), \(\cdots\), then the following \(E_{l-r}\) can measure the error. We call it ``left-right golden error".

\[\begin{split}
E_{l-r} = &\sum_{i=0}^{l}\sum_{j=0}^{m-2}\Bigg| \frac{x_{im+j+1} \!-\!x_{im+j}}{x_{im+j+2} \!-\! x_{im+j}} \!+\! (-1)^j\varphi \!+\! sign\big(\!-\! 1 \!-\! (-1)^j\big) \Bigg| +\!\\
&\sum_{j=m(l+1)}^{n-2}\Bigg| \frac{x_{j+1} - x_j}{x_{j+2} - x_j} + (-1)^j\varphi + sign\big(-1 - (-1)^j\big) \Bigg|,
\end{split}\]
where
\[sign(x) = \left\{
\begin{aligned}
&1, x>0,\\
&0, x=0,\\
&-1, x<0,
\end{aligned}
\right.\]
is the sign function.

If we hope the ratios are \(1 / \varphi\), \(\varphi\), \(1 / \varphi\), \(\varphi\), \(\cdots\), then we can easily construct a ``right-left golden error" \(E_{r-l}\) by replacing the exponent \(j\) by \(j+1\) in the above formula.

\end{enumerate}

The above four kinds of errors will increase with the increase of \(n\). In order to compare the golden degree of different step functions, we can compute the average error by multiplying the error by \[\frac{1}{(m-1)(l+1)+\big(n-1-m(l+1)\big)_+}\] to eliminate the influence of \(n\), where
\[(x)_+ = \left\{
\begin{aligned}
&x, x>0,\\
&0, x\le0.
\end{aligned}
\right.\]

The above errors are convenient for numerical comparison, but when we want to minimize them, we often need to solve their derivatives, in which case we can replace the above absolute value operation with the square operation for convenience.

For the golden step interpolation, suppose the original interpolation nodes \(A_0(x_0,y_0)\), \(A_1(x_1,y_1)\), \(\cdots\), \(A_n(x_n,y_n)\) are transformed into \(A'_0(x'_0,y'_0)\), \(A'_1(x'_1,y'_1)\), \(\cdots\), \(A'_k(x'_k,y'_k)\), then the interpolation problem can be turned into the following optimization problem.
\begin{align*}
&\min\quad E(x'_0,x'_1,\cdots,x'_k)\\
& \begin{array}{lll}
s.t.&\text{the final function passes through the original nodes;}\\
&\text{other constraints on }x'_0,x'_1,\cdots,x'_k,\\
\end{array}
\end{align*}
where \(E\) represents one of the above five errors: \(E_{left}\), \(E_{right}\), \(E_{mixed}\), \(E_{l-r}\) and \(E_{r-l}\).

When \(m<n\), it is a local golden method, in which the golden optimization only happens in every group consisting of \(m+1\) nodes; When \(m=n\), it becomes a global golden method, in which the optimization target is to make the ratio of every two adjacent line segments satisfies \(\varphi\) or \(1/\varphi\) in the whole function graph.

It is easy to work out that the golden extension step interpolation proposed in Section~\ref{sec:stepGE} is the optimum solution of the following optimization problem when \(m=2\) and \(k=2n\).
\begin{align*}
&\min\quad E_{left}(x'_0,x'_1,\cdots,x'_k)\\
& \begin{array}{lll}
s.t.&x'_0=x_0;\\
&x_0<x'_1<x_1;\\
&x'_2=x_1;\\
&x_1<x'_3<x_2;\\
&x'_4=x_2;\\
&x_2<x'_5<x_3;\\
&\cdots\\
&x_{n-1}<x'_{k-1}<x_n;\\
&x'_k=x_n.\\
\end{array}
\end{align*}

And the golden equal number step interpolation proposed in Section~\ref{sec:stepG} is the optimum solution of the following optimization problem when \(m=2\) and \(k=n\).
\begin{align*}
&\min\quad E_{left}(x'_0,x'_1,\cdots,x'_k)\\
& \begin{array}{lll}
s.t.&x'_0=x_0;\\
&x_0<x'_1 \le x_1;\\
&x'_2=x_2;\\
&x_2<x'_3 \le x_3;\\
&x'_4=x_4;\\
&x_4<x'_5 \le x_5;\\
&\cdots\\
&x_{2[0.5n]-2}<x'_{2[0.5n]-1} \le x_{2[0.5n]-1};\\
&\text{If \(n\) is odd, add a constraint: }x'_{k-1}=x_{n-1}; \\
&x'_k=x_n.\\
\end{array}
\end{align*}


\section{Golden piecewise linear interpolation}\label{sec:linear}
The step function is discontinuous. In this section, we go more complex to the simplest continuous interpolation function: the piecewise linear interpolation, and we discuss how to apply the golden section principle to it. The graph of a piecewise linear function is composed of connected line segments end to end. If we use the traditional piecewise linear interpolation method, the interpolation data nodes are exactly the ends of the line segments, so it is convenient to change the function graph by transforming the nodes. Considering that, we add a golden transform to the traditional method, struggling to make the interpolation graph look more beautiful.


\subsection{Golden cuspidal hill}\label{subsec:cuspidal}
When thinking how to apply the golden section principle to the piecewise linear interpolation, we may firstly call to mind the golden triangles that are universally acknowledged the most beautiful among all triangles. The ratio of the lengths of a golden triangle's sides conforms to the golden section ratio. Therefore, we naturally consider making the lengths' ratio of two adjacent line segments of the piecewise line satisfy the golden section ratio. In our study, we have indeed done this first of all. But unfortunately, we find this strategy is bad and the interpolation graph looks even uglier. Thinking the reason carefully, it may be that compared with the triangle, the piecewise line consisting of two segments lacks of an edge, which leads to essential difference between the piecewise line and the triangle. So the golden principle in triangles is not applicable to piecewise lines.

Now that the idea inspired by the golden triangles fails, we propose another strategy in this paper. In our strategy, we first define the concept of the golden cuspidal hill as follows.

\begin{figure}[H]\centering
  \includegraphics[scale=.6]{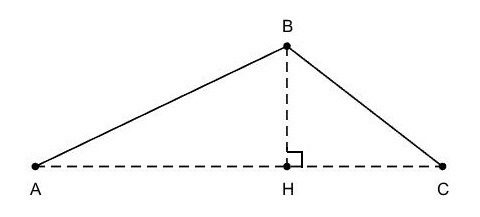}
  \caption{ Golden cuspidal hill. }
  \label{fig:cuspidalHill}
\end{figure}

\begin{definition}
See~\reffig{fig:cuspidalHill}. A piecewise line \(ABC\) consists of two segments: \(AB\) and \(BC\). Then \(ABC\) is called cuspidal hill, and the point \(B\) is called the hilltop. Connect \(AC\) and construct a perpendicular line \(BH\) intersecting \(AC\) at the point \(H\). If \(H\) is a golden section point of \(AC\), then we call \(ABC\) a golden cuspidal hill and call \(B\) the golden hilltop. If \(H\) is the left(right) golden point, then \(ABC\) is called the left(right) golden cuspidal hill and \(B\) is called the left(right) golden hilltop.
\end{definition}

We argue that golden cuspidal hills have beautiful elements among all piecewise lines with two segments. It is inspired by a golden application about the location of a host: It is suggested that a host should stand at the golden section point of the stage, because standing in the middle of the stage seems dull and standing too far from the middle seems disharmony. The location of the golden section point is considered natural and aesthetic. The golden cuspidal hills refer to the similar principle: The hilltop \(B\) is a break point of the graph. When this break happens in the golden section point of \(AC\), the cuspidal hill will seem natural, harmonious and aesthetic. The basic idea of our golden piecewise linear interpolation is to make the interpolation nodes connected by golden cuspidal hills.


\subsection{Golden extension piecewise linear interpolation}\label{subsec:gel}
We add an extension transform to the piecewise linear interpolation, which adds an extra data node between each two adjacent interpolation nodes, and finally make every two adjacent interpolation nodes connected by a golden cuspidal hill. The specific interpolation process includes two steps as follows, and we call this method the golden extension piecewise linear interpolation.

\begin{enumerate}[step 1.]
  \item Transform \({A_0}\), \({A_1}\), \(\cdots\), \({A_n}\) into \({{A}_{0}}\), \({{A}_{0.5}}\), \({{A}_{1}}\), \({{A}_{1.5}}\), \(\cdots\) , \({{A}_{n-0.5}}\), \({{A}_{n}}\) as follows:

      Add an extra data node \({{A}_{i+0.5}}({{x}_{i+0.5}},{{y}_{i+0.5}})\) between each two adjacent interpolation nodes \({{A}_{i}},{{A}_{i+1}}\)\((i=0,1,\cdots,n-1)\), making the piecewise line connecting \(A_i,A_{i+0.5},A_{i+1}\) become a golden cuspidal hill. That is to say, we should make the node \({{A}_{i+0.5}}\) become the golden hilltop of \({{A}_{i}}{{A}_{i+0.5}}{{A}_{i+1}}\). See~\reffig{fig:solveCuspidalHillE}. There are the nodes \({{A}_{i}}\) and \({{A}_{i+1}}\). Connect \({{A}_{i}}{{A}_{i+1}}\). Take the right golden section point \(H\) of the line segment \({{A}_{i}}{{A}_{i+1}}\). Construct \({{A}_{i+0.5}}H\)\(\perp\)\({{A}_{i}}{{A}_{i+1}}\) and make \(|{{A}_{i+0.5}}H|=q|{{A}_{i}}{{A}_{i+1}}|\) where \(q=0.2\) whose value is given by us. Then \({{A}_{i+0.5}}\) is the added node that we want. The coordinates of  \(A_{i+0.5}\) can be solved by elementary mathematics and we put the solving process into Appendix~\ref{appendix:gel}. Here we just list the important results.

      \begin{figure}[H]\centering
      \includegraphics[scale=.4]{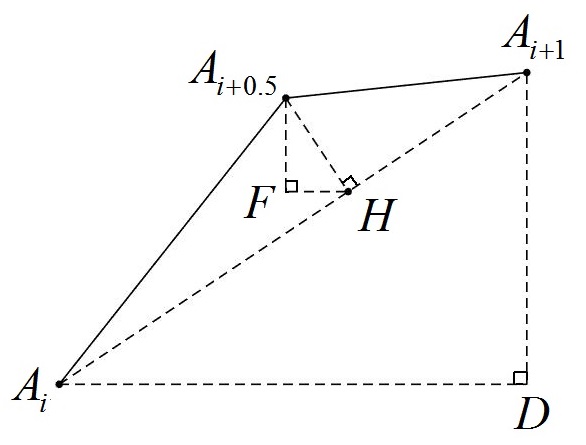}
      \caption{ Add a node \({{A}_{i+0.5}}\) between \({{A}_{i}}\) and \({{A}_{i+1}}\). }
      \label{fig:solveCuspidalHillE}
      \end{figure}

      The slope of \({{A}_{i}}{{A}_{i+1}}\) is \[k = \frac{{{y_{i + 1}} - {y_i}}}{{{x_{i + 1}} - {x_i}}}.\]

      The x-coordinate and the y-coordinate of \({{A}_{i+0.5}}\) can be solved by the following formulas:
      \begin{equation}
           {x_{i + 0.5}} = \left\{ \begin{aligned}
&{x_i} + ({x_{i + 1}} - {x_i})\varphi {\rm{ + }}q({y_{i + 1}} - {y_i}),\left| k \right| \ge 1,\\
&{x_i} + ({x_{i + 1}} - {x_i})\varphi  - q({y_{i + 1}} - {y_i}),\left| k \right| < 1,
\end{aligned} \right.
           \label{con:lgx-coordinate}
      \end{equation}
      \begin{equation}
      {y_{i + 0.5}} = \left\{ \begin{aligned}
&{y_i} + ({y_{i + 1}} - {y_i})\varphi  - q({x_{i + 1}} - {x_i}),\left| k \right| \ge 1,\\
&{y_i} + ({y_{i + 1}} - {y_i})\varphi  + q({x_{i + 1}} - {x_i}),\left| k \right| < 1,
\end{aligned} \right.
           \label{con:y-coordinate}
      \end{equation}
      where \[q=\frac{{|{A_{i + 0.5}}H|}}{{|{A_i}{A_{i + 1}|}}}=0.2\] whose value is given by us.

      Finally we should limit the scope of \({{x}_{i+0.5}}\) in order to avoid \({x_{i + 0.5}}\) is not in \(({x_i},{x_{i + 1}})\) or too close to the endpoint of \(({x_i},{x_{i + 1}})\). We let \(t=0.5(1-\varphi )(x_{i+1}-x_{i})\). If the \({{x}_{i+0.5}}\) calculated according to~\eqref{con:lgx-coordinate} and~\eqref{con:y-coordinate} is not in the interval \([{{x}_{i}}+t,{{x}_{i+1}}-t]\), then we revise the value of \(q\) by letting \[q=\frac{t}{\left| {{y}_{i+1}}-{{y}_{i}} \right|},\]and recalculate \({{x}_{i+0.5}}\) and \({{y}_{i+0.5}}\) according to~\eqref{con:lgx-coordinate} and~\eqref{con:y-coordinate}.

    \item Get the final interpolation function \[{p_{le}}(x) = \frac{{x - {x_{0.5(i + 1)}}}}{{{x_{0.5i}} - {x_{0.5(i + 1)}}}}{y_{0.5i}} + \frac{{x - {x_{0.5i}}}}{{{x_{0.5(i + 1)}} - {x_{0.5i}}}}{y_{0.5(i + 1)}},\]where \(x \in [{x_{0.5i}},{x_{0.5(i + 1)}}],i = 0,1, \cdots,2n - 1.\)
\end{enumerate}


\subsection{Golden equal number piecewise linear interpolation}\label{subsec:gl}
By adding data nodes, the golden extension piecewise linear interpolation makes any two adjacent interpolation data nodes connected by a golden cuspidal hill. However, compared with the direct piecewise linear interpolation, due to adding data nodes, this method makes the non-differentiable points almost double in the interpolation function. In the function graph, the number of line segments increases and the length of each line segment becomes shorter. But sometimes, we want to make the graph look more beautiful on condition that the interpolation function has the same number of non-differentiable points as the direct piecewise linear interpolation function. Therefore, we also propose a golden equal number transform, in which three adjacent interpolation data nodes are extracted in turn and the data node in the middle of the three is transformed, which is aimed to make the final interpolation function graph not only pass through the original interpolation nodes but also consist of more golden cuspidal hills. The specific interpolation process includes two steps as follows, and we call this method the golden equal number piecewise linear interpolation.

\begin{enumerate}
\renewcommand{\labelenumi}{step \theenumi.}
  \item  Transform \({{A}_{0}},{{A}_{1}},\cdots,{{A}_{n}}\) into \({{A}_{0}},{{{A}'}_{1}},{{A}_{2}},{{{A}'}_{3}},\cdots,{{A}_{n}}\) as follows:

If \(n<2\), the data nodes are not changed; that is, the transformed data nodes are the same as the original data nodes.

If \(n\ge 2\), take out the data nodes \(A_{2i-1}(x_{2i-1},y_{2i-1}),i=1,2,\cdots,[0.5n]\) whose sequence numbers are odd numbers among the \(n\) data nodes, in which \([0.5n]\) represents the integer part of \(0.5n\). Transform \(A_{2i-1}(x_{2i-1},y_{2i-1})\) into \(A'_{2i-1}(x'_{2i-1},y'_{2i-1})\). The process of solving \(A'_{2i-1}(x'_{2i-1},y'_{2i-1})\) is illustrated in~\reffig{fig:solveCuspidalHill}. There are data nodes \(A_{2i-2}\), \(A_{2i-1}\) and \(A_{2i}\). Connect \(A_{2i-2}A_{2i}\). Construct \(A_{2i-1}P\)\(\perp\)\(A_{2i-2}A_{2i}\) where \(P\) is the foot point. If \(P\) is not on the line segment \(A_{2i-2}A_{2i}\) or is a endpoint of the line segment \(A_{2i-2}A_{2i}\), then we let \(A'_{2i-1}\) the same as \({{A}_{2i-1}}\); otherwise, we find the golden section point \(H\) of \(A_{2i-2}A_{2i}\) where \(H\) is the the golden section point that is closer to \(P\). Construct \(A'_{2i-1}H\)\(\perp\)\(A_{2i-2}A_{2i}\) where \(H\) is the foot point. If the x-coordinate of \(H\) is greater than the x-coordinate of \(P\), then extend the line segment \(A_{2i-2}A_{2i-1}\), making the extended line and \(A'_{2i-1}H\) intersect at the point \(A'_{2i-1}\); if the x-coordinate of \(H\) is equal to or smaller than the x-coordinate of \(P\), then extend the line segment \(A_{2i}A_{2i-1}\), making the extended line and \(A'_{2i-1}H\) intersect at the point \(A'_{2i-1}\). Then, the piecewise line \(A_{2i-2}A'_{2i-1}A_{2i}\) not only passes through the original nodes \(A_{2i-2}\), \(A_{2i-1}\) and \(A_{2i}\), but also is a golden cuspidal hill. However, the x-coordinate of \(A'_{2i-1}\) may not be in the interval \((x_{2i-2},x_{2i})\), in which condition the x-coordinates of the transformed data nodes may not be monotonically increasing so we can't construct interpolation function for them. When this happens, we make \(A'_{2i-1}\) the same as \(A_{2i-1}\); that is, we don't change \({{A}_{2i-1}}\). Therefore, this transform just tries to do better but can't guarantee that the piecewise line \(A_{2i-2}A'_{2i-1}A_{2i}\) becomes a golden cuspidal hill. The specific calculation steps and formulas to solve \(A'_{2i-1}\) are as follows.

\begin{figure}[H]\centering
  \includegraphics[scale=.4]{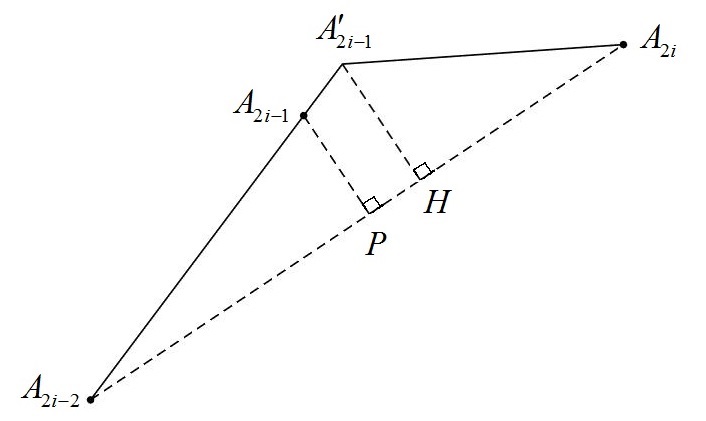}
  \caption{ Transform the node \({{A}_{2i-1}}\) into the node \({{{A}'}_{2i-1}}\). }
  \label{fig:solveCuspidalHill}
\end{figure}

\begin{enumerate}
\renewcommand{\labelenumii}{step \theenumii. }
  \item Solve \(P\). Let \(k\) stand for the slope of \({{A}_{2i-2}}{{A}_{2i}}\). Then \[k = \frac{{{y_{2i}} - {y_{2i - 2}}}}{{{x_{2i}} - {x_{2i - 2}}}}.\] Using the knowledge of elementary mathematics, we can calculate that the coordinates of \(P\)  are
\[\left\{
\begin{aligned}
&{x_p} = \frac{{k({y_{2i - 1}} - {y_{2i - 2}}) + {k^2}{x_{2i - 2}} + {x_{2i - 1}}}}{{1 + {k^2}}},\\
&{y_p} = \frac{{k({x_{2i - 1}} - {x_{2i - 2}}) + {k^2}{y_{2i - 1}} + {y_{2i - 2}}}}{{1 + {k^2}}}.
\end{aligned}
\right.\]
If \({{x}_{p}}\notin ({{x}_{2i-2}},{{x}_{2i}})\), then make \({{{A}'}_{2i-1}}\) the same as \({{A}_{2i-1}}\); that is, \[\left\{ \begin{aligned}
&{{x'}_{2i - 1}} = {x_{2i - 1}},\\
&{{y'}_{2i - 1}} = {y_{2i - 1}}.
\end{aligned} \right.\] If \({{x}_{p}}\in ({{x}_{2i-2}},{{x}_{2i}})\), then continue the following steps.

  \item Solve \(H\). The x-coordinate of the midpoint of the line segment \({{A}_{2i-2}}{{A}_{2i}}\) is \[{x_z} = \frac{{{x_{2i - 2}} + {x_{2i}}}}{2}.\] If \({{x}_{p}}>{{x}_{z}}\), then the coordinates of \(H\) are \[\left\{ \begin{aligned}
&{x_h} = {x_{2i-2}}{\rm{ + }}(x_{2i} - x_{2i-2})\varphi, \\
&{y_h} = {y_{2i-2}}{\rm{ + }}(y_{2i} - y_{2i-2})\varphi;
\end{aligned} \right.\] if \({{x}_{p}}\le {{x}_{z}}\), then the coordinates of \(H\) are \[\left\{ \begin{aligned}
&{x_h} = {x_{2i}} - (x_{2i} - x_{2i-2})\varphi, \\
&{y_h} = {y_{2i}} - (y_{2i} - y_{2i-2})\varphi.
\end{aligned} \right.\]

  \item Solve \({{{A}'}_{2i-1}}\). Let \[c = \left\{
\begin{aligned}
&\frac{{{y_{2i - 1}} - {y_{2i - 2}}}}{{{x_{2i - 1}} - {x_{2i - 2}}}},{x_h} > {x_p},\\
&\frac{{{y_{2i - 1}} - {y_{2i}}}}{{{x_{2i - 1}} - {x_{2i}}}},{x_h} \le {x_p}.
\end{aligned}
\right.\] Then the coordinates of \({{{A}'}_{2i-1}}\) are \[\left\{
\begin{aligned}
&{x'_{2i - 1}} = \frac{{x_h + k(c{x_{2i - 1}} + y_h - y_{2i - 1})}}{{1 + ck}},\\
&{y'_{2i - 1}} = \frac{{y_{2i - 1} + c(k{y_h} + x_h - x_{2i - 1})}}{{1 + ck}}.
\end{aligned}
\right.\]

  \item  Check and decide the value. If \({{{x}'}_{2i-1}}\notin ({{x}_{2i-2}},{{x}_{2i}})\), then make \({{{A}'}_{2i-1}}\) the same as \({{A}_{2i-1}}\); that is, \[\left\{ \begin{aligned}
&{{x'}_{2i - 1}} = {x_{2i - 1}},\\
&{{y'}_{2i - 1}} = {y_{2i - 1}}.
\end{aligned} \right.\]
\end{enumerate}

\item Mark \(A_0,A'_1,A_2,A'_3,\cdots,A_n\) as \(A'_0(x'_0,y'_0)\), \(A'_1(x'_1,y'_1)\), \(A'_2(x'_2,y'_2)\), \(A'_3(x'_3,y'_3)\), \(\cdots\) , \(A'_n(x'_n,y'_n)\),  and then get the final interpolation function\[p_l(x) = \frac{x - x'_{i + 1}}{x'_i - x'_{i + 1}}y'_i + \frac{x - x'_i}{x'_{i + 1} - x'_i}y'_{i + 1},\]where \(x \in [x'_i,x'_{i + 1}],i = 0,1, \cdots,n-1.\)
\end{enumerate}


\subsection{Examples and applications}
\begin{figure*}[htbp] \centering
\subfigure[] { \label{fig:line}
\includegraphics[width=0.65\columnwidth]{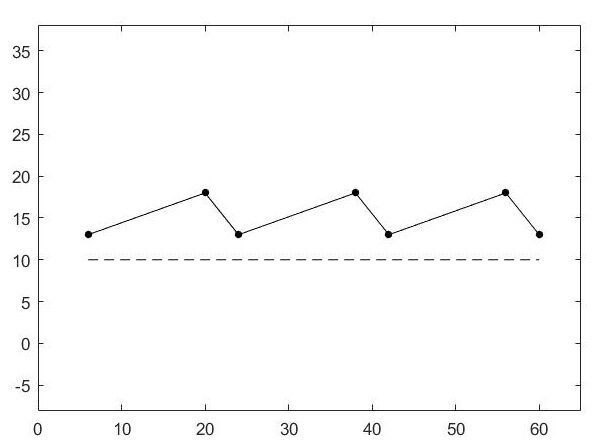}
}
\subfigure[] { \label{fig:lineGE}
\includegraphics[width=0.65\columnwidth]{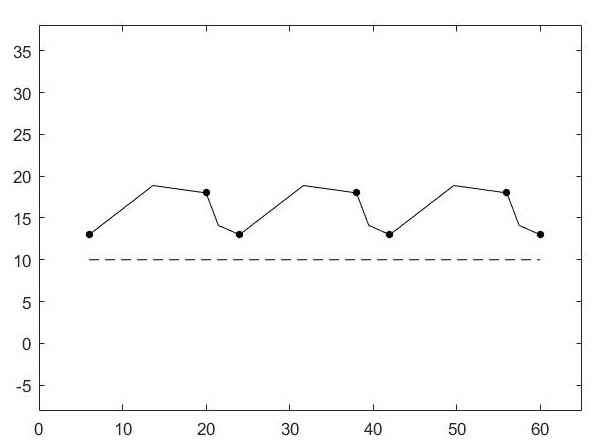}
}
\subfigure[] { \label{fig:lineG}
\includegraphics[width=0.65\columnwidth]{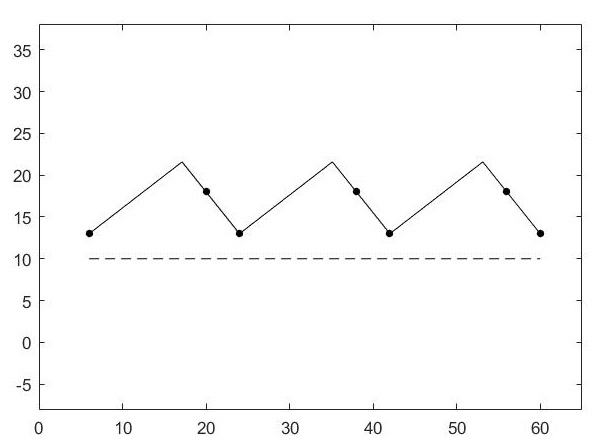}
}
\\
\subfigure[] { \label{fig:lineApp}
\includegraphics[width=0.65\columnwidth]{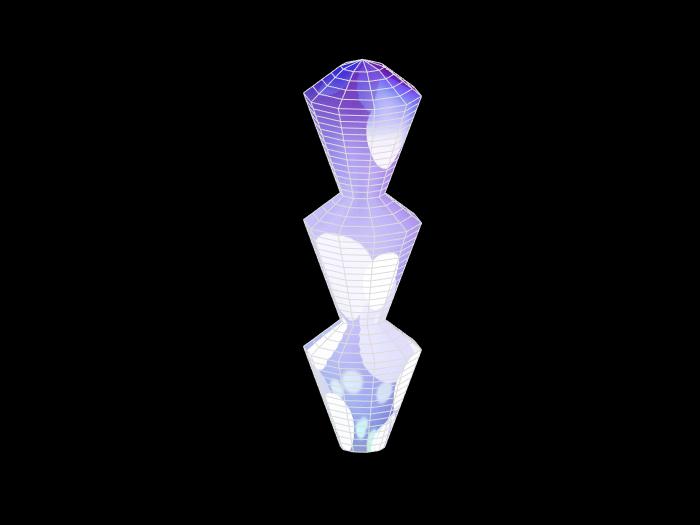}
}
\subfigure[] { \label{fig:lineGEApp}
\includegraphics[width=0.65\columnwidth]{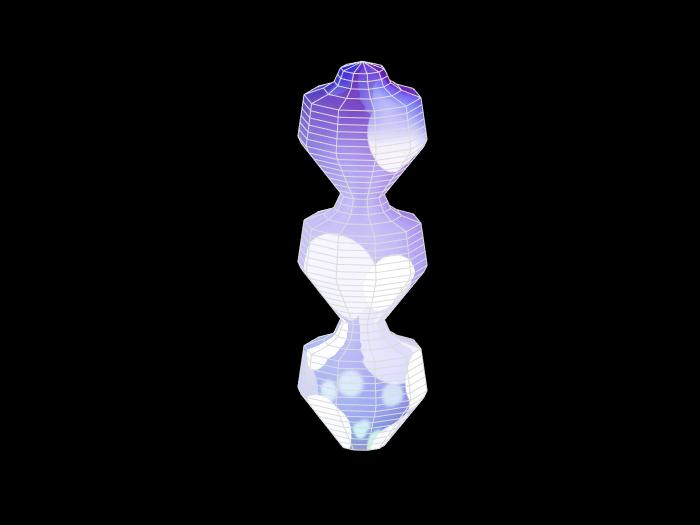}
}
\subfigure[] { \label{fig:lineGApp}
\includegraphics[width=0.65\columnwidth]{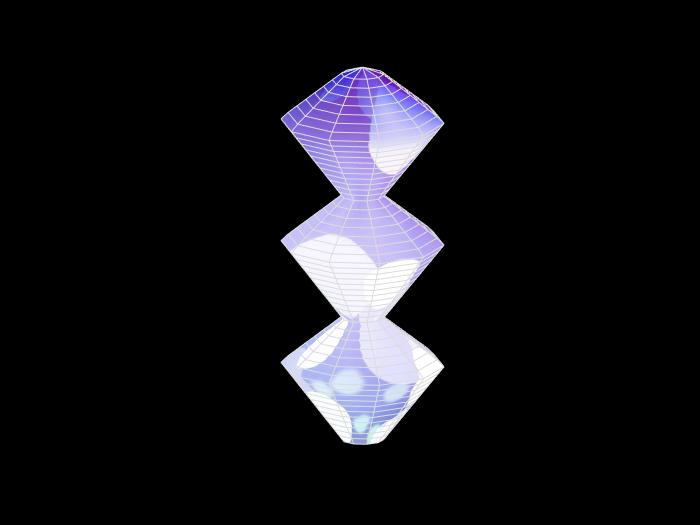}
}
\caption{  Comparative examples and applications: (a) the graph of the traditional piecewise linear interpolation; (b) the graph of the golden extension piecewise linear interpolation; (c) the graph of the golden equal number piecewise linear interpolation; (d) the 3D landscape light generated by (a); (e) the 3D landscape light generated by (b); (f) the 3D landscape light generated by (c).  }
\label{fig:gline}
\end{figure*}

We demonstrate comparative examples and applications about the shape design of landscape lights: Given design nodes \(C_0(6,13)\), \(C_1(20,18)\), \(C_2(24,13)\), \(C_3(38,18)\), \(C_4(42,13)\), \(C_5(56,18)\), \(C_6(60,13)\), construct the profiles of the landscape lights by doing the traditional piecewise linear interpolation, the golden extension piecewise linear interpolation and the golden equal number piecewise linear interpolation respectively for them. The graphs of these three interpolation functions are shown in~\reffig{fig:line},~\reffig{fig:lineGE} and~\reffig{fig:lineG}, which are the three profiles constructed by the three different methods. It should be noted that the unit of the x-axis should be equal to the unit of the y-axis when drawing, so as to avoid the distortion of the graph in the visual effect. Rotate the profiles around the line \(y = 10\), and then the 3D models of the landscape lights can be obtained, which are shown in~\reffig{fig:lineApp},~\reffig{fig:lineGEApp} and~\reffig{fig:lineGApp}. Compared with~\reffig{fig:lineApp}, ~\reffig{fig:lineGEApp} and~\reffig{fig:lineGApp} integrate the golden section into the geometric features, which makes the designed model look more pleasant.

\subsection{Go further to explore criteria}\label{subsec:glCriteria}
We can explore criteria for the golden piecewise linear interpolation in the similar way of exploring the golden step interpolation criteria.

Suppose the \(n+1\) nodes \(A_0(x_0,y_0)\), \(A_1(x_1,y_1)\), \(\cdots\) , \(A_n(x_n,y_n)\) are directly input to the piecewise linear interpolation. For a node \(A_{i+1}(i=0,1,\cdots,n-2)\), it is the hilltop of the cuspidal hill \(A_iA_{i+1}A_{i+2}\). Construct \(A_{i+1}C\) \(\perp\) \(A_iA_{i+2}\) where \(C(x_c,y_c)\) is the foot point.

\begin{figure}[H]\centering
  \includegraphics[scale=.45]{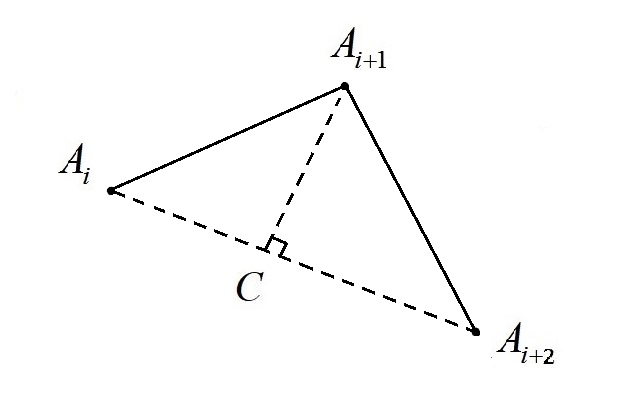}
  \caption{ A cuspidal hill \(A_iA_{i+1}A_{i+2}\). }
  \label{fig:aCuspidalHill}
\end{figure}

Then
\[x_c = \frac{k(y_{i+1} - y_i) + {k^2}x_i + x_{i+1}}{1 + k^2}, \text{ where } k = \frac{y_{i+2} - y_i}{x_{i+2} - x_i}.\]

Get the ratio
\[\begin{split}
t_{i+1}&=\frac{x_c-x_i}{x_{i+2}-x_i}\\
&=\frac{(y_{i+2}-y_i)(y_{i+1}-y_i)+(x_{i+2}-x_i)(x_{i+1}-x_i)}{(x_{i+2}-x_i)^2+(y_{i+2}-y_i)^2}.
\end{split}\]

Raplace the
\[\frac{x_{im+j+1} - x_{im+j}}{x_{im+j+2} - x_{im+j}} \text{ and } \frac{x_{j+1} - x_j}{x_{j+2} - x_j}\]
in Section~\ref{subsec:criteriaStep} with \(t_{im+j+1}\) and \(t_{j+1}\), and then we can get the errors \(E_{left}\), \(E_{right}\), \(E_{mixed}\), \(E_{l-r}\) and \(E_{r-l}\) for the the golden piecewise linear interpolation with a value of \(m\) given. What's different from the step interpolation is that the ratios \(t_{im+j+1}\) and \(t_{j+1}\) may be negative when \(\angle A_{i+1}A_iA_{i+2}\) is an obtuse angle, but it doesn't matter for the evaluation of the errors. The meanings of these kinds of errors are as follows.

\begin{enumerate}[(1)]

\item The left(right) golden error \(E_{left}\)(\(E_{right}\))

Every \(m+1\) nodes forms a group. If we hope every cuspidal hill in each group is a left(right) golden cuspidal hill, then the error \(E_{left}\)(\(E_{right}\)) can measure the distance between the present graph and the goal.

\item The mixed golden error \(E_{mixed}\)

If we hope every cuspidal hill in each group is a golden cuspidal hill, no matter what kind of golden cuspidal hill it is, then the error \(E_{mixed}\) can be a measure.

\item The alternate golden error \(E_{l-r}\) and \(E_{r-l}\)

If we hope the cuspidal hills in each group are ``the left golden cuspidal hill, the right golden cuspidal hill, the left golden cuspidal hill, the right golden cuspidal hill...'', then the ``left-right golden error" \(E_{l-r}\) can be a measure. Similarly, the meaning of the ``right-left golden error" \(E_{r-l}\) is easy to known.

\end{enumerate}

As for the average error, the square error and the relevant optimization problems, they are similar to those in Section~\ref{subsec:criteriaStep}. However, for the golden piecewise linear interpolation, the optimization problems and their constraints are more complex.

Suppose the original interpolation nodes \(A_0(x_0,y_0)\), \(A_1(x_1,y_1)\), \(\cdots\), \(A_n(x_n,y_n)\) are transformed into \(A'_0(x'_0,y'_0)\), \(A'_1(x'_1,y'_1)\), \(\cdots\), \(A'_k(x'_k,y'_k)\).

The golden extension piecewise linear interpolation proposed in Section~\ref{subsec:gel} is one of the optimum solutions of the following optimization problem when \(m=2\) and \(k=2n\).
\begin{align*}
&\min\quad E_{right}(x'_0,x'_1,\cdots,x'_k,y'_0,y'_1,\cdots,y'_k)\\
& \begin{array}{lll}
s.t.&x'_0=x_0 \text{ and } y'_0=y_0;\\
&x_0+0.5(1-\varphi )(x_1-x_0) \le x'_1 \le x_1-0.5(1-\varphi )(x_1-x_0);\\
&x'_2=x_1 \text{ and } y'_2=y_1;\\
&x_1+0.5(1-\varphi )(x_2-x_1) \le x'_3 \le x_2-0.5(1-\varphi )(x_2-x_1);\\
&x'_4=x_2 \text{ and } y'_4=y_2;\\
&x_2+0.5(1-\varphi )(x_3-x_2) \le x'_5 \le x_3-0.5(1-\varphi )(x_3-x_2);\\
&\cdots\\
&x_{n\!-\!1}\!+\!0.5(1\!-\!\varphi )(x_n\!-\!x_{n\!-\!1}) \le x'_{k-1} \le x_n\!-\!0.5(1\!-\!\varphi )(x_n\!-\!x_{n\!-\!1});\\
&x'_k=x_n \text{ and } x'_k=x_n.\\
\end{array}
\end{align*}

The above optimization problem has countless optimal solutions. The setting of \(q\) and the choice of the above node or the below node according to the slope \(k\) in Section~\ref{subsec:gel} are to select a solution from all the optimal solutions.

The golden equal number piecewise linear interpolation proposed in Section~\ref{subsec:gl} is related to the following optimization problem when \(m=2\) and \(k=n\). But it isn't the optimal solution and needs to be improved.
\begin{align*}
&\min\quad E_{mixed}(x'_0,x'_1,\cdots,x'_k,y'_0,y'_1,\cdots,y'_k)\\
& \begin{array}{lll}
s.t.&x'_0=x_0 \text{ and } y'_0=y_0;\\
&\bigg\{(x'_1,y'_1)\Big|x_0 < x'_1 \le x_1,\displaystyle\frac{y'_1-y_1}{x'_1-x_1}=\frac{y_2-y_1}{x_2-x_1}\bigg\}\bigcup\\
&\bigg\{(x'_1,y'_1)\Big|x_1 < x'_1 < x_2,\displaystyle\frac{y'_1-y_1}{x'_1-x_1}=\frac{y_0-y_1}{x_0-x_1}\bigg\};\\
&x'_2=x_2 \text{ and } y'_2=y_2;\\
&\bigg\{(x'_3,y'_3)\Big|x_2 < x'_3 \le x_3,\displaystyle\frac{y'_3-y_3}{x'_3-x_3}=\frac{y_4-y_3}{x_4-x_3}\bigg\}\bigcup\\
&\bigg\{(x'_3,y'_3)\Big|x_3 < x'_3 < x_4,\displaystyle\frac{y'_3-y_3}{x'_3-x_3}=\frac{y_2-y_3}{x_2-x_3}\bigg\};\\
&x'_4=x_4 \text{ and } y'_4=y_4;\\
&\bigg\{(x'_5,y'_5)\Big|x_4 < x'_5 \le x_5,\displaystyle\frac{y'_5-y_5}{x'_5-x_5}=\frac{y_6-y_5}{x_6-x_5}\bigg\}\bigcup\\
&\bigg\{(x'_5,y'_5)\Big|x_5 < x'_5 < x_6,\displaystyle\frac{y'_5-y_5}{x'_5-x_5}=\frac{y_4-y_5}{x_4-x_5}\bigg\};\\
&\cdots\\
&\bigg\{(x'_{2[0.5n]-1},y'_{2[0.5n]-1})\Big|x_{2[0.5n]-2} < x'_{2[0.5n]-1} \le x_{2[0.5n]-1}, \\ &\displaystyle\frac{y'_{2[0.5n]-1}-y_{2[0.5n]-1}}{x'_{2[0.5n]-1}-x_{2[0.5n]-1}} = \frac{y_{2[0.5n]}-y_{2[0.5n]-1}}{x_{2[0.5n]}-x_{2[0.5n]-1}}\bigg\}\bigcup\\
&\bigg\{(x'_{2[0.5n]-1},y'_{2[0.5n]-1})\Big|x_{2[0.5n]-1} < x'_{2[0.5n]-1} < x_{2[0.5n]}, \\ &\displaystyle\frac{y'_{2[0.5n]-1}-y_{2[0.5n]-1}}{x'_{2[0.5n]-1}-x_{2[0.5n]-1}}=\frac{y_{2[0.5n]-2}-y_{2[0.5n]-1}}{x_{2[0.5n]-2}-x_{2[0.5n]-1}}\bigg\};\\
&\text{If \(n\) is odd, add constraints: \(x'_{k-1}=x_{n-1}\) and \(y'_{k-1}=y_{n-1}\);} \\
&x'_k=x_n \text{ and } y'_k=y_n.\\
\end{array}
\end{align*}

The above optimization problem is rough. It may have more than one optimal solutions, so we need further strategies to select one. The error of the optimal solution may be very large, so some constraints on errors need be further added(So is it in Section~\ref{subsec:gl}). All like these need to be further considered in practical methods.


\section{Golden curve interpolation}\label{sec:curve}
In previous sections, we have discussed the discontinuous step function and the \(C^0\) smooth piecewise linear function. In this section, we go further to the \(C^1\) smooth function. We'll discuss how to apply the golden section principle to a smooth interpolation curve with a continuous first-order derivative for aesthetic. Spline functions are made up of piecewise polynomials with a certain degree of smoothness. They have both smoothness and flexibility. Among them, the quadratic spline has a continuous first-order derivative, and it also has a good convexity-preserving. Lots of work studies the shape-preserving interpolation curve based on the quadratic spline~\cite{r9shapepreserving,r10shapepreserving,r11shapepreserving}. That is because there will be no excess inflection points when we interpolate with a quadratic spline. For these reasons, we choose the quadratic spline interpolation as a basis to control shape conveniently, and we add a golden extension transform to it, aiming to integrate the golden aesthetic into the interpolation curve.
\subsection{Golden domed hill}
We have defined the cuspidal hill in section~\ref{subsec:cuspidal}. Similarly, in this section, we propose the concept of the golden domed hill for introducing our golden curve interpolation.

\begin{figure}[H]\centering
  \includegraphics[scale=.6]{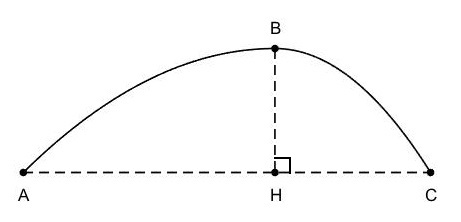}
  \caption{ Golden domed hill. }
  \label{fig:domedHill}
\end{figure}

\begin{definition}
See~\reffig{fig:domedHill}. In the plane right-angle coordinate system, there is a curve with \(A(x_a,{y_a})\) and \(C(x_c,y_c)\) as the end points and \(x_a<x_c\). Suppose that the functional expression of the curve is \(f(x)\). \(f(x)\) has a continuous first-order derivative \(f'(x)\) and has no inflection points in the interval \((x_a,x_c)\). According to the Lagrange mean value theorem, there exists \(x_b\in (x_a,x_c)\) which can make \[f'(x_b)=\frac{y_a-y_c}{x_a-x_c};\] that is, there exists a node \(B(x_b,f(x_b))\) on the curve where the tangent slope at the node \(B\) equals the slope of the line \(AC\). Construct \(BH\)\(\perp\)\(AC\) where \(H\) is the foot point. If the node \(H\) is a golden section point of the line segment \(AC\), then the curve \(ABC\) is called a golden domed hill or a golden domed hill curve. And the node \(B\) is called the golden hilltop of the curve \(ABC\). If \(H\) is the left(right) golden point, then \(ABC\) is called the left(right) golden domed hill and \(B\) is called the left(right) golden hilltop.
\end{definition}

We argue that golden domed hill curves are the aesthetic curves. The reason is similar to the golden cuspidal hill. Therefore, for the golden curve interpolation, our basic idea is to make the function graph contain more golden domed hills.

\subsection{Golden extension curve interpolation}\label{sec:gecurve}
Given interpolation data nodes \({{A}_{0}}({{x}_{0}},{{y}_{0}})\), \({{A}_{1}}({{x}_{1}},{{y}_{1}})\), \(\cdots\) , \({{A}_{n}}({{x}_{n}},{{y}_{n}})\) and the derivative \({{k}_{0}}\) at the node \({{A}_{0}}\), then their quadratic spline interpolation function is uniquely determined. If a spline knot \({{x}_{i+0.5}}\) is added between every two adjacent data nodes \({{A}_{i}}, {{A}_{i+1}}\)(\(i=0,1,\cdots,n-1\)) but the value of the function at \({{x}_{i+0.5}}\) is not given, then the quadratic spline interpolation function is not unique and increases \(n\) degrees of freedom. The basic idea of our golden extension curve interpolation is to take advantage of the \(n\) degrees of freedom to make the interpolation data nodes connected by more golden domed hills.

To unify all of our golden methods in one framework: the ``node transform + traditional interpolation'' pattern, so that we can edit the shape of the function graph by editing the data nodes, we do the golden extension curve interpolation according to the following two steps.

\begin{enumerate}[step 1.]

\item Transform \({A_0}\), \({A_1}\), \(\cdots\), \({A_n}\) into \(A_0\), \((A_{0.5})\), \(A_1\), \((A_{1.5})\), \(\cdots\), \((A_{n-0.5})\), \(A_n\) as follows. Here the nodes in brackets ``()'' may exist and may not exist, which is determined by the following specific computation.

    Suppose we add a node \(A_{i+0.5}(x_{i+0.5},y_{i+0.5})\) between the node \(A_i\) and \(A_{i+1}\)(\(i=0,1,\cdots,n-1\)), and we try to make the section of the curve in the interval \([{{x}_{i}},{{x}_{i+1}}]\) become a golden domed hill curve and make the data node \(A_{i+0.5}\) become the golden hilltop of this section of the curve. Here we take the right domed hill for example, and the left domed hill in a similar way.

    Let \(H\) be the right golden point of the line segment \(A_iA_{i+1}\). Then the coordinates of \(H\) are
    \[\left\{ \begin{aligned}
    &{x_h} = {x_i}{\rm{ + }}({x_{i + 1}} - {x_i})\varphi, \\
    &{y_h} = {y_i}{\rm{ + }}({y_{i + 1}} - {y_i})\varphi.
    \end{aligned} \right.\]

    We have got the transformed nodes \(A_0\), \((A_{0.5})\), \(A_1\), \((A_{1.5})\), \(\cdots\), \(A_i\) in the previous iterations, and for convenience we mark them as \(B_0(x'_0,y'_0)\), \(B_1(x'_1,y'_1)\), \(\cdots\), \(B_d(x'_d,y'_d)\), where \(d+1\) is the actual number of the nodes \(A_0\), \((A_{0.5})\), \(A_1\), \((A_{1.5})\), \(\cdots\), \(A_i\).

    Suppose the expression of the quadratic spline interpolation function is \(p(x)\). Referring to the Section~\ref{sec:interpolation}, we can get
    \[\begin{split}
    p'(x_{i+0.5})=&2\Bigg(\frac{y_{i+0.5}-y_i}{x_{i+0.5}-x_i} + \sum_{j=0}^{d-1}(-1)^{d-j}\frac{y'_{j+1}-y'_j}{x'_{j+1}-x'_j}\Bigg) + \\
    &(-1)^{d+1}{k_0}.
    \end{split}\]

    If \(A_{i+0.5}\) is the right golden hilltop, we can list the following equations:

    When \(y_{i + 1} = y_i\),

    \begin{equation}
    \left\{ \begin{aligned}
    &p'(x_{i+0.5}) = \frac{y_{i+1}-y_i}{x_{i+1}-x_i},\\
    &x_{i+0.5} = x_h.\\
    \end{aligned} \right.
    \label{con:gc0}
    \end{equation}

    When \(y_{i + 1} \ne y_i\),

    \begin{equation}
    \left\{ \begin{aligned}
    &p'(x_{i+0.5}) = \frac{y_{i+1}-y_i}{x_{i+1}-x_i},\\
    &\frac{y_{i+0.5}-y_h}{x_{i+0.5}-x_h} \centerdot \frac{y_{i+1}-y_i}{x_{i+1}-x_i}= -1.\\
    \end{aligned} \right.
    \label{con:gc1}
    \end{equation}

    Mark
    \[t=\frac{1}{2}\Bigg(\frac{y_{i+1}-y_i}{x_{i+1}-x_i}-(-1)^{d+1}k_0\Bigg)-\sum_{j=0}^{d-1}(-1)^{d-j}\frac{y'_{j+1}-y'_j}{x'_{j+1}-x'_j}.\]

    Then the equations~\eqref{con:gc0} above can be solved according to the following formulae~\eqref{con:gc0Solve}.

    \begin{equation}
    \left\{ \begin{aligned}
    &x_{i+0.5} = x_h,\\
    &y_{i+0.5} = y_i+t(x_{i+0.5}-x_i).\\
    \end{aligned} \right.
    \label{con:gc0Solve}
    \end{equation}

    And the equations~\eqref{con:gc1} above can be solved according to the following formulae~\eqref{con:gc1Solve}.

    \begin{equation}
    \left\{ \begin{aligned}
    &x_{i+0.5} = \frac{(y_{i+1}-y_i)(y_h-y_i+tx_i)+x_h(x_{i+1}-x_i)}{t(y_{i+1}-y_i)+x_{i+1}-x_i},\\
    &y_{i+0.5} = y_i+t(x_{i+0.5}-x_i).\\
    \end{aligned} \right.
    \label{con:gc1Solve}
    \end{equation}

    We can observe that the formulae~\eqref{con:gc0Solve} and~\eqref{con:gc1Solve} can be unified to the one formulae~\eqref{con:gc1Solve}.

    Finally, we should judge whether \(x_{i+0.5}\) is valid. If \(x_{i+0.5}\in (x_i,x_{i+1})\), then add the node \(A_{i+0.5}\)(that is, set \((A_{i+0.5})\) as \(A_{i+0.5}\)); if \(x_{i+0.5}\notin ({{x}_{i}},{{x}_{i+1}})\), then don't add the node \(A_{i+0.5}\)(that is, set \((A_{i+0.5})\) to be empty).

\item Mark the nodes \(A_0\), \((A_{0.5})\), \(A_1\), \((A_{1.5})\), \(\cdots\), \((A_{n-0.5})\), \(A_n\) as \(B_0(x'_0,y'_0)\), \(B_1(x'_1,y'_1)\), \(\cdots\), \(B_d(x'_d,y'_d)\), where \(d+1\) is the actual number of the transformed nodes. Then get the final interpolation function
    \[\begin{split}
    p(x) = &{\left(\frac{x - x'_{i + 1}}{x'_i - x'_{i + 1}}\right)}^{2}y'_i + {\left(\frac{x - x'_i}{{x'_{i + 1} - x'_i}}\right)}^{2}y'_{i + 1} + \\
    &\frac{(x - x'_i)(x - x'_{i + 1})}{x'_i - x'_{i + 1}}\Bigg[2\sum_{j=0}^{i-1}(-1)^{i-j+1}\frac{y'_{j + 1} - y'_j}{x'_{j + 1} - x'_j} + \\
    &{(-1)^i}{k_0} - \frac{2y'_i}{x'_i - x'_{i + 1}}\Bigg],
    \end{split} \]
    where \(x \in [x'_i,x'_{i + 1}],i = 0,1, \cdots,d - 1\).

\end{enumerate}

There is something to explain here. The golden curve interpolation method just tries to do better but can't guarantee that the curve between two adjacent interpolation data nodes becomes a golden domed hill. There exist some cases in which the constructed interpolation function is the same as the traditional quadratic spline interpolation function. Error control isn't done in the method, so there are some cases where errors are unbearably large. We don't always want every two adjacent nodes to be connected by a golden domed hill, and a golden domed hill is not better than a traditional curve all the time. Therefore, in practical applications we can delete some added nodes to edit the curve and make the curves in some intervals become the traditional quadratic spline interpolation curves according to practical needs.
\subsection{Examples and applications}
To compare the golden method and the traditional method, we demonstrate two examples and applications about shape design.

\begin{figure*}[htbp] \centering
\subfigure[] { \label{fig:VaseCurve}
\includegraphics[width=0.55\columnwidth]{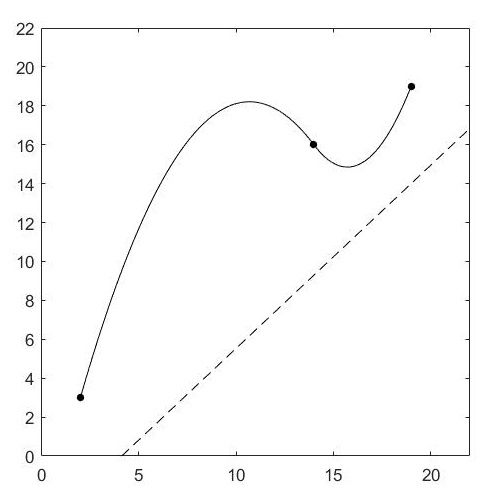}
}
\hspace{44pt}
\subfigure[] { \label{fig:VaseCurveGE}
\includegraphics[width=0.55\columnwidth]{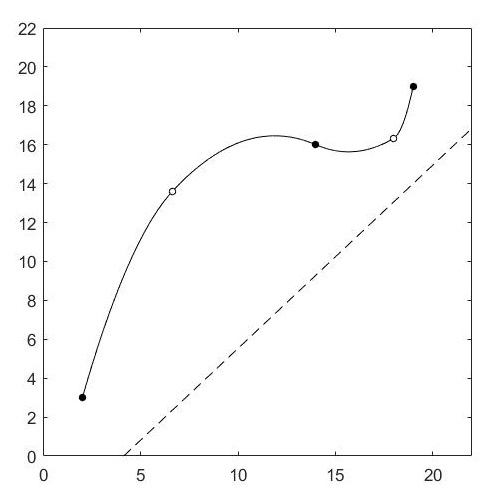}
}
\\
\subfigure[] { \label{fig:VaseCurveApp}
\includegraphics[width=0.45\columnwidth]{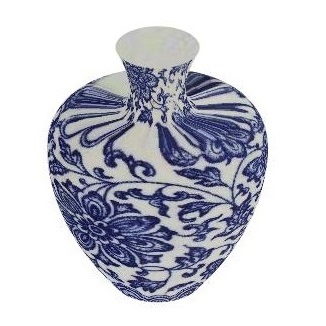}
}
\hspace{72pt}
\subfigure[] { \label{fig:VaseCurveGEApp}
\includegraphics[width=0.45\columnwidth]{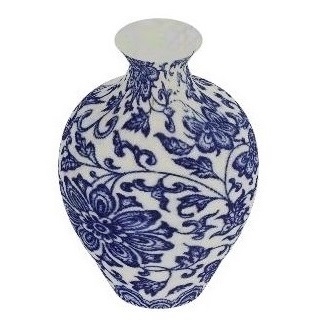}
}
\caption{ Comparative examples and applications: (a) the graph of the traditional quadratic spline interpolation; (b) the graph of the golden extension curve interpolation; (c) the 3D vase model generated by (a); (d) the 3D vase model generated by (b). }
\label{fig:gcurve}
\end{figure*}

Interpolation curves are often used to construct the rotational surface models for designing glass cups, vases and other objects. Here we show the shape design of a vase: Given data nodes \(D_0(2,3)\), \(D_1(14,16)\), \(D_2(19,19)\) and the derivative \(k_0=3.5\) at the node \(D_0\), do the traditional quadratic spline interpolation and the golden extension curve interpolation respectively for them. The graphs of these two interpolation functions are shown in~\reffig{fig:VaseCurve} and~\reffig{fig:VaseCurveGE}. In the graph, the solid dots are the interpolation data nodes and the hollow dots are the golden hilltops. Note that the unit of the x-axis should be equal to the unit of the y-axis when drawing the graph. Rotate the two curves around the line \(16x-17y-66=0\), and then the two vase models can be obtained, which are shown in~\reffig{fig:VaseCurveApp} and~\reffig{fig:VaseCurveGEApp}. We can see the vase generated by the golden method is more beautiful because the hilltops of the two pieces of curves are golden hilltops.

\begin{figure*}[htbp] \centering
\subfigure[] { \label{fig:BedCurve}
\includegraphics[width=0.75\columnwidth]{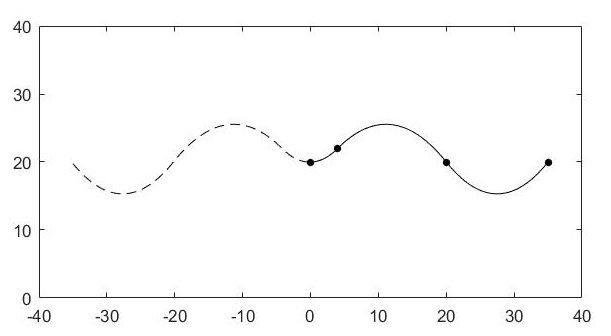}
}
\subfigure[] { \label{fig:BedCurveGE}
\includegraphics[width=0.75\columnwidth]{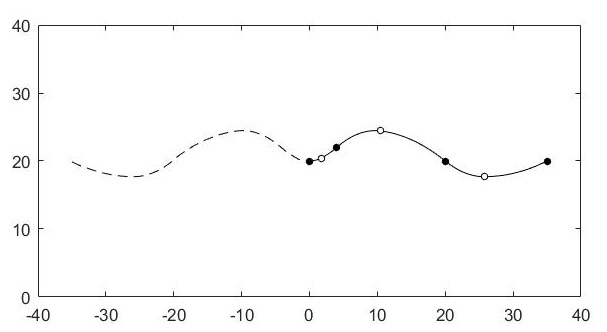}
}
\\
\hspace{2pt}
\subfigure[] { \label{fig:BedCurveApp}
\includegraphics[width=0.7\columnwidth]{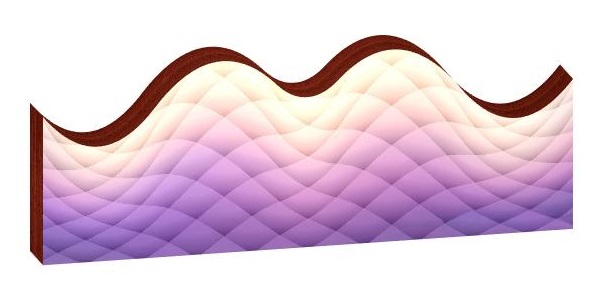}
}
\hspace{15pt}
\subfigure[] { \label{fig:BedCurveGEApp}
\includegraphics[width=0.7\columnwidth]{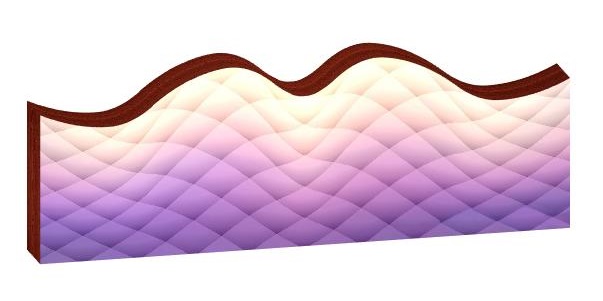}
}
\caption{ Comparative examples and applications: (a) the graph of the traditional quadratic spline interpolation; (b) the graph of the golden extension curve interpolation; (c) the 3D headboard model generated by (a); (d) the 3D headboard model generated by (b). }
\label{fig:gcurve}
\end{figure*}

Another application is the design of a headboard shape. For nodes \(E_0(0,20)\), \(E_1(4,22)\), \(E_2(20,20)\), \(E_3(35,20)\) and the derivative \(k_0=0\) at \(E_0\), their function graphs of the traditional quadratic spline interpolation and the golden extension curve interpolation are shown in~\reffig{fig:BedCurve} and~\reffig{fig:BedCurveGE}. In this application we take the left golden section points in the golden method. Make a curve symmetrical with the interpolation curve with respect to the line \(x=0\) in each graph, and then the corresponding headboard shapes are designed. The designed 3D headboards are shown in~\reffig{fig:BedCurveApp} and~\reffig{fig:BedCurveGEApp}. We can see the second headboard which is designed by the golden method looks a little more beautiful because it is composed of a number of golden domed hills.

The golden curve interpolation integrates the aesthetic characteristics of the golden section into the geometric curve, which is beneficial for designing more beautiful models.

\subsection{Go further to explore criteria}
Based on the idea of the golden domed hill, how do we measure the golden degree of a curve?

\begin{figure}[H]\centering
  \includegraphics[scale=.55]{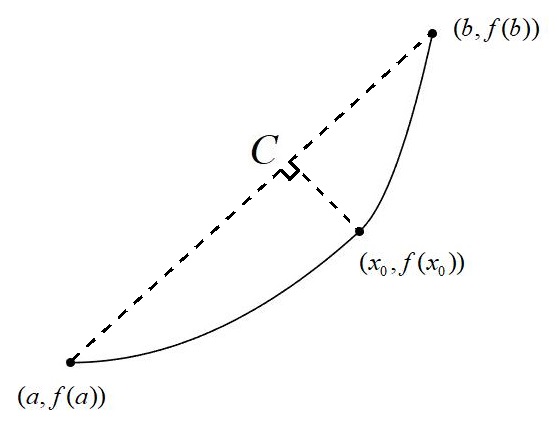}
  \caption{ A domed hill. }
  \label{fig:aDomedHill}
\end{figure}

Consider a curve \(f(x)\) in the interval \([a,b]\). Suppose \(f(x)\) has a continuous first-order derivative \(f'(x)\) and has no inflection points in the interval \((a,b)\). See~\reffig{fig:aDomedHill}. It is a domed hill. How much far away is it from a golden domed hill?

Suppose \(x_0\) is the root of the equation \[f'(x)=\frac{f(a)-f(b)}{a-b}.\]

Then \((x_0,f(x_0))\) is the hilltop of the curve. Construct a perpendicular from the hilltop like~\reffig{fig:aDomedHill}. We can easily know the x-coordinate of the foot point \(C\) is
\[x_c = \frac{k(f(x_0) - a) + {k^2}a + x_0}{1 + k^2}, \text{ where } k = \frac{f(a)-f(b)}{a-b}.\]

It is similar to Section~\ref{subsec:glCriteria}. Get the ratio
\[\begin{split}
t&=\frac{x_c-a}{b-a}\\
&=\frac{(f(b)-f(a))(f(x_0)-f(a))+(b-a)(x_0-a)}{(b-a)^2+(f(b)-f(a))^2}.
\end{split}\]

Then the absolute value
\[|t-\varphi|, |t-(1-\varphi)|, \text{ and } \left\{ \begin{aligned}
    &|t-\varphi|, t \ge 0.5,\\
    &|t-(1-\varphi)|, t<0.5,
    \end{aligned} \right.\]
can be used to measure how far away the curve is form the right golden domed hill, the left golden domed hill and any golden domed hill respectively.

We have discussed how to measure a domed hill. However, an interpolation curve often contains several domed hills. How do we measure such a curve?

Suppose a curve \(F(x)\) is defined in the interval \([c,d]\). And there are  \(z\) points \(e_1,e_2,\cdots,e_z\) partitioning \([c,d]\) into \(z+1\) intervals. In each of the \(z+1\) intervals, the corresponding curve of \(F(x)\) is a domed hill. Then the golden degree of the whole curve \(F(x)\) can be measured by \[E=\sum_{i=1}^{z+1}E_i,\] where \(E_i\) represents a certain golden error(among the above three) of the ith domed hill. And the average golden error of \(F(x)\) is \[E=\frac{1}{z+1}\sum_{i=1}^{z+1}E_i.\]

We have discussed how to measure the golden degree of a curve and we can observe: The golden measure of a curve is related to how to partition the  definition domain of the curve function. For the same curve, the golden errors may vary with different partitions. So the partition is important and it indicates which curve segments we want to become golden domed hills. For the golden extension curve interpolation proposed in Section~\ref{sec:gecurve}, the interpolation data nodes play two roles: (1) interpolation constrains; (2) partition the definition domain. And we can understand that we choose to use the quadratic spline is to make the curve segments in the partitioned intervals become domed hills without inflection points.

After the problem of golden measure, then we will discuss how to convert the golden curve interpolation problem into a more general optimization problem.

Reconsider the golden extension curve interpolation in Section~\ref{sec:gecurve}. We abstract the problem into the following form.

Give interpolation conditions \(A_0(x_0,y_0)\), \(A_1(x_1,y_1)\), \(\cdots\) , \(A_n(x_n,y_n)\).

Give partition points \(x_1, x_2, \cdots , x_{n-1}\) in the definition domain \([x_0,x_n]\).

Add nodes \(A_{0.5}(x_{0.5},y_{0.5})\), \(A_{1.5}(x_{1.5},y_{1.5})\), \(\cdots\) , \(A_{n-0.5}(x_{n-0.5},y_{n-0.5})\). They are the unknowns.

Then the quadratic spline interpolation function \(p(x)\) can be constructed for the nodes \(A_0\), \(A_{0.5}\), \(A_1\), \(A_{1.5}\), \(\cdots\), \(A_{n-0.5}\), \(A_n\). And the golden error \(E\) of \(p(x)\) can be calculated as mentioned above. Then the golden extension curve interpolation in Section~\ref{sec:gecurve} is related to the following optimization problem but it isn't the optimum solution.

\begin{align*}
&\min\quad E_{right}(x_{0.5},x_{1.5},\cdots,x_{n-0.5},y_{0.5},y_{1.5},\cdots,y_{n-0.5})\\
& \begin{array}{lll}
s.t.&A_{0.5}, A_{1.5}, \cdots, A_{n-0.5} \text{ are golden hilltops};\\
&x_{0.5}, x_{1.5}, \cdots, x_{n-0.5} \text{ are in the intervals } (x_0,x_1),(x_1,x_2), \\
&\cdots, (x_{n-1},x_n) \text{ respectively};\\
\end{array}
\end{align*}
where we put the unknowns \(x_{0.5}\), \(x_{1.5}\), \(\cdots\), \(x_{n-0.5}\), \(y_{0.5}\), \(y_{1.5}\), \(\cdots\), \(y_{n-0.5}\) into the ``()'' of the ``\(E_{right}()\)''.

The above optimization problem requires that the added nodes should be golden hilltops. A more general problem is removing this condition as follows.

\begin{align*}
&\min\quad E(x_{0.5},x_{1.5},\cdots,x_{n-0.5},y_{0.5},y_{1.5},\cdots,y_{n-0.5})\\
& \begin{array}{lll}
s.t.&x_{0.5}, x_{1.5}, \cdots, x_{n-0.5} \text{ are in the intervals } (x_0,x_1),(x_1,x_2), \\
&\cdots, (x_{n-1},x_n) \text{ respectively}.\\
\end{array}
\end{align*}

But solving this problem is more complicated.

Similarly, other optimization problems of the golden curve interpolation can be established. For example, here the interpolation data nodes play a role in partitioning the definition domain, but we can give another partition to specially express which curve segments we want to become golden domed hills. For another example, here we only propose a kind of node transform, but more various node transforms are worth exploring.

\section{Conclusion and prospect}\label{sec:conclusion}
For the classic aesthetic interpolation problem, this paper has proposed a novel solving thought that is applying the golden section principle. But how should we apply the golden section to interpolation methods, making the interpolation graph not only show the personality characteristics to the given nodes, but also be integrated into the aesthetic characteristics of the golden section? To answer this question, this paper starts from the simplest case, and gradually goes to more complex, which gives the examples of applying the golden section to the interpolation of degree 0, 1 and 2. And in the last two kinds of examples, we have proposed the novel and interesting concepts of the golden cuspidal hill and the golden domed hill. The related ideas, methods, criteria, comparative examples and applications are also presented. The entirely new idea for the classic interpolation problem is promising. This paper is to do the initial exploration and provide the reference for further researches and better methods.

At the end of this paper, we also point out what needs to be improved in our work and what is worth exploring in the future: (1) If the philosophy of pursuing interpolation which meet ``golden" principle would be accompanied by more analyses, validation and user study, the novel thought will be more convincing. (2) The parametric curve interpolation is more widely used, so it is very meaningful to explore how to apply the golden section principle to these parametric curves. (3) This paper only presents one-dimensional golden interpolation methods, but the higher dimensional golden interpolation is worth exploring in the future. (4) For applying the golden section principle to interpolation, we only initially explore three examples, but more applications and better methods need to be further explored.

\section*{Acknowledgments}
We would like to thank anonymous reviewers for their insightful comments and suggestions which greatly improve the quality of this paper.
We thank Professor Liu Ligang from University of Science and Technology of China for his concern and guidance for this paper.
We thank Zhao Wang for helping to modify the format of this paper.
We are grateful for the Numerical Methods course of North China University of Science and Technology which is a national level excellent course.
This work was partially supported by the National Nature Science Foundation of China (51674121, 61702184) and the Scientific and Technological Project for the Returned Overseas Scholars in HeBei (C2015005014).

\titleformat{\section}{\normalfont\normalsize\bfseries}{Appendix~\Alph{section}.}{5pt}{\normalsize}
\begin{appendices}
\section{The solving process of \(A_{i+0.5}\) in Section~\ref{subsec:gel}}\label{appendix:gel}
To solve the \({{A}_{i+0.5}}\) in section~\ref{subsec:gel}, we add some auxiliary lines in~\reffig{fig:solveCuspidalHillE}: construct \({{A}_{i+0.5}}F\)\(\perp\)\(FH\) where \(F\) is the foot point; make \({{A}_{i}}D\)\(\perp\)\(D{{A}_{i+1}}\) where \(D\) is the foot point.

\(H\) is the golden section point that is closer to \({{A}_{i+1}}\), so the coordinates of \(H\) are \[\left\{
\begin{aligned}
&{x_h} = {x_i} + ({x_{i + 1}} - {x_i})\varphi,\\
&{y_h} = {y_i} + ({y_{i + 1}} - {y_i})\varphi.
\end{aligned}
\right.\]The slope of \({{A}_{i}}{{A}_{i+1}}\) is \[k = \frac{{{y_{i + 1}} - {y_i}}}{{{x_{i + 1}} - {x_i}}}.\]

      When \(k=0\), the coordinates of \({{A}_{i+0.5}}\) can be calculated by the formula \[\left\{
\begin{aligned}
&{x_{i + 0.5}} = {x_i} + ({x_{i + 1}} - {x_i})\varphi,\\
&{y_{i + 0.5}} = {y_i} + q({x_{i + 1}} - {x_i}).
\end{aligned}
\right.\]

      When \(k\ne 0\), it is obvious that \(\Delta\)\({{A}_{i+0.5}}FH\)\(\sim\)\(\Delta\)\({{A}_{i}}D{{A}_{i+1}}\). Then \[\frac{|{{A_{i + 0.5}}F}|}{{|{A_i}D|}} = \frac{{|FH|}}{{|D{A_{i + 1}|}}} = \frac{{|{A_{i + 0.5}}H|}}{{|{A_i}{A_{i + 1}|}}} = q.\]So \[|{{A}_{i+0.5}}F|=q|{{A}_{i}}D|=q({{x}_{i+1}}-{{x}_{i}}),\] and \[|FH|=q|D{{A}_{i+1}|}=q\left| {{y}_{i+1}}-{{y}_{i}} \right|.\] We give the regulation that when \(\left| k \right|>1\), the node \({{A}_{i+0.5}}\) is below \({{A}_{i}}{{A}_{i+1}}\) and when \(\left| k \right|<1\), the node \({{A}_{i+0.5}}\) is above \({{A}_{i}}{{A}_{i+1}}\).

      Therefore, when \(\left| k \right|>1\), the x-coordinate of \({{A}_{i+0.5}}\) is \[{{x}_{i+0.5}}={{x}_{h}}+sign(k)\cdot |FH|={{x}_{i}}+({{x}_{i+1}}-{{x}_{i}})\varphi +q({{y}_{i+1}}-{{y}_{i}}),\] and the y-coordinate of \({{A}_{i+0.5}}\) is \[{{y}_{i+0.5}}={{y}_{h}}-|{{A}_{i+0.5}}F|={{y}_{i}}+({{y}_{i+1}}-{{y}_{i}})\varphi -q({{x}_{i+1}}-{{x}_{i}});\] when \(\left| k \right|<1\), the x-coordinate of \({{A}_{i+0.5}}\) is \[{{x}_{i+0.5}}={{x}_{h}}-|FH|={{x}_{i}}+({{x}_{i+1}}-{{x}_{i}})\varphi -q\left| {{y}_{i+1}}-{{y}_{i}} \right|,\] and the y-coordinate of \({{A}_{i+0.5}}\) is \[{{y}_{i+0.5}}={{y}_{h}}+|{{A}_{i+0.5}}F|={{y}_{i}}+({{y}_{i+1}}-{{y}_{i}})\varphi +q({{x}_{i+1}}-{{x}_{i}}).\]

      In summary, the x-coordinate formula of \({{A}_{i+0.5}}\) is

      \[
           {x_{i + 0.5}} = \left\{ \begin{aligned}
&{x_i} + ({x_{i + 1}} - {x_i})\varphi {\rm{ + }}q({y_{i + 1}} - {y_i}),\left| k \right| \ge 1,\\
&{x_i} + ({x_{i + 1}} - {x_i})\varphi  - q({y_{i + 1}} - {y_i}),\left| k \right| < 1,
\end{aligned} \right.
      \]

      and the y-coordinate formula of \({{A}_{i+0.5}}\) is

      \[
      {y_{i + 0.5}} = \left\{ \begin{aligned}
&{y_i} + ({y_{i + 1}} - {y_i})\varphi  - q({x_{i + 1}} - {x_i}),\left| k \right| \ge 1,\\
&{y_i} + ({y_{i + 1}} - {y_i})\varphi  + q({x_{i + 1}} - {x_i}),\left| k \right| < 1.
\end{aligned} \right.
      \]

\section{Algorithms of our golden methods}\label{appendix:algorithm}

\begin{algorithm}[H]
\caption{Golden extension step interpolation} 
\label{alg:gestep}
\hspace*{0.02in} {\bf Input:} 
\({A_0}({x_0},{y_0})\), \({A_1}({x_1},{y_1})\), \(\cdots\) , \({A_n}({x_n},{y_n})\), a given \(x \in [x_0,x_n)\)\\
\hspace*{0.02in} {\bf Output:} 
the value of \({{p}_{se}}(x)\)
\begin{algorithmic}[1]
\State // Do the golden extension transform 
\For{\(i=0:n-1\)} 
¡¡¡¡\State Create a new node \({{A}_{i+0.5}}({{x}_{i+0.5}},{{y}_{i+0.5}})\);
¡¡¡¡\State \({{x}_{i+0.5}} \gets {{x}_{i+1}}-({{x}_{i+1}}-{{x}_{i}})*\varphi\);
¡¡¡¡\If{\({y_i} \ne {y_{i + 1}}\)} 
¡¡¡¡¡¡¡¡\State \({{y}_{i+0.5}} \gets {{y}_{i+1}}-({{y}_{i+1}}-{{y}_{i}})*\varphi\);
¡¡¡¡\Else
¡¡¡¡¡¡¡¡\State \({{y}_{i+0.5}} \gets L\);
¡¡¡¡\EndIf
\EndFor
\State // Do the traditional interpolation for the transformed nodes
\For{\(i=0:2n-1\)} 
¡¡¡¡\If{\(x\in [{{x}_{0.5i}},{{x}_{0.5(i+1)}})\)} 
¡¡¡¡¡¡¡¡\State \Return \({{y}_{0.5i}}\);
¡¡¡¡\EndIf
\EndFor
\end{algorithmic}
\end{algorithm}

\begin{algorithm}[H]
\caption{Golden equal number step interpolation}
\label{alg:gstep}
\hspace*{0.02in} {\bf Input:}
\({A_0}({x_0},{y_0})\), \({A_1}({x_1},{y_1})\), \(\cdots\) , \({A_n}({x_n},{y_n})\), a given \(x \in [x_0,x_n)\)\\
\hspace*{0.02in} {\bf Output:}
the value of \({{p}_{s}}(x)\)
\begin{algorithmic}[1]
\State // Do the golden equal number transform
\If{\(n\ge 2\)}
¡¡¡¡\For{\(i = 1:Integer(0.5n)\)}
¡¡¡¡¡¡¡¡\State \({x_g} \gets {{x}_{2i}}-({{x}_{2i}}-{{x}_{2i-2}})*\varphi \);
¡¡¡¡¡¡¡¡\If{\({x_g} < {x_{2i - 1}}\)}
¡¡¡¡¡¡¡¡¡¡¡¡\State \({x_{2i - 1}} \gets {x_g}\);
¡¡¡¡¡¡¡¡\EndIf
¡¡¡¡\EndFor
\EndIf
\State // Do the traditional interpolation for the transformed nodes
\For{\(i=0:n-1\)}
¡¡¡¡\If{\(x\in [x_i,x_{i+1})\)}
¡¡¡¡¡¡¡¡\State \Return \({{y}_{i}}\);
¡¡¡¡\EndIf
\EndFor
\end{algorithmic}
\end{algorithm}

\begin{algorithm}[H]
\caption{Golden extension piecewise linear interpolation}
\label{alg:geline}
\hspace*{0.02in} {\bf Input:}
\({A_0}({x_0},{y_0})\), \({A_1}({x_1},{y_1})\), \(\cdots\) , \({A_n}({x_n},{y_n})\), a given \(x \in [x_0,x_n]\)\\
\hspace*{0.02in} {\bf Output:}
the value of \({{p}_{le}}(x)\)
\begin{algorithmic}[1]
\State // Do the golden extension transform
\For{\(i=0:n-1\)}
¡¡¡¡\State Create a new node \({{A}_{i+0.5}}({{x}_{i+0.5}},{{y}_{i+0.5}})\);
¡¡¡¡\State \(k \gets (y_{i+1}-y_i)/(x_{i+1}-x_i)\);
¡¡¡¡\State \(q \gets 0.2\);
¡¡¡¡\If{\(|k| \ge 1\)}
¡¡¡¡¡¡¡¡\State \({x_{i+0.5}} \gets x_i + (x_{i + 1} - x_i)*\varphi + q*(y_{i + 1} - y_i)\);
¡¡¡¡¡¡¡¡\State \({y_{i+0.5}} \gets y_i + (y_{i + 1} - y_i)*\varphi - q*(x_{i + 1} - x_i)\);
¡¡¡¡\Else
¡¡¡¡¡¡¡¡\State \({x_{i+0.5}} \gets x_i + (x_{i + 1} - x_i)*\varphi - q*(y_{i + 1} - y_i)\);
¡¡¡¡¡¡¡¡\State \({y_{i+0.5}} \gets y_i + (y_{i + 1} - y_i)*\varphi + q*(x_{i + 1} - x_i)\);
¡¡¡¡\EndIf
¡¡¡¡\State \(t \gets 0.5*(1-\varphi )*(x_{i+1}-x_i)\)
¡¡¡¡\If{\(x_{i+0.5} \notin [x_i+t,x_{i+1}-t]\)}
¡¡¡¡¡¡¡¡\State \(q \gets t/|y_{i+1}-y_i|\);
¡¡¡¡¡¡¡¡\State Do the previous ``if...else...'' codes again here;
¡¡¡¡\EndIf
\EndFor
\State // Do the traditional interpolation for the transformed nodes
\For{\(i=0:2n-1\)}
¡¡¡¡\If{\(x \in [{x_{0.5i}},{x_{0.5(i + 1)}}]\)}
¡¡¡¡¡¡¡¡\State \Return \((x-x_{0.5(i+1)})/(x_{0.5i}-x_{0.5(i+1)})*y_{0.5i}+(x-x_{0.5i})/(x_{0.5(i+1)}-x_{0.5i})*y_{0.5(i+1)}\);
¡¡¡¡\EndIf
\EndFor
\end{algorithmic}
\end{algorithm}

\begin{algorithm}[H]
\caption{Golden equal number piecewise linear interpolation} 
\label{alg:gline}
\hspace*{0.02in} {\bf Input:}
\({A_0}({x_0},{y_0})\), \({A_1}({x_1},{y_1})\), \(\cdots\) , \({A_n}({x_n},{y_n})\), a given \(x \in [{x_i},{x_{i + 1}}]\)\\
\hspace*{0.02in} {\bf Output:}
the value of \({p_l}(x)\)
\begin{algorithmic}[1]
\State // Do the golden extension transform
\If{\(n\ge 2\)}
¡¡¡¡\For{\(i = 1:Integer(0.5n)\)}
¡¡¡¡¡¡¡¡\State // Step a. Solve \(P\)
¡¡¡¡¡¡¡¡\State \(k \gets (y_{2i}-y_{2i-2})/(x_{2i}-x_{2i-2})\);
¡¡¡¡¡¡¡¡\State \(x_p \gets (k*(y_{2i-1}-y_{2i-2})+k*k*x_{2i-2}+x_{2i-1})/(1+k*k)\);
¡¡¡¡¡¡¡¡\State \(y_p \gets (k*(x_{2i-1}-x_{2i-2})+k*k*y_{2i-1}+y_{2i-2})/(1+k*k)\);
¡¡¡¡¡¡¡¡\If{\(x_p \in (x_{2i-2},x_{2i})\)}
¡¡¡¡¡¡¡¡¡¡¡¡\State // Step b. Solve \(H\)
¡¡¡¡¡¡¡¡¡¡¡¡\State \(x_z \gets (x_{2i-2}+x_{2i})/2\);
¡¡¡¡¡¡¡¡¡¡¡¡\If{\(x_p > x_z\)}
¡¡¡¡¡¡¡¡¡¡¡¡¡¡¡¡\State \({x_h} \gets {x_{2i-2}}+(x_{2i} - x_{2i-2})*\varphi\);
¡¡¡¡¡¡¡¡¡¡¡¡¡¡¡¡\State \({y_h} \gets {y_{2i-2}}+(y_{2i} - y_{2i-2})*\varphi\);
¡¡¡¡¡¡¡¡¡¡¡¡\Else
¡¡¡¡¡¡¡¡¡¡¡¡¡¡¡¡\State \({x_h} \gets {x_{2i}} - (x_{2i} - x_{2i-2})*\varphi\);
¡¡¡¡¡¡¡¡¡¡¡¡¡¡¡¡\State \({y_h} \gets {y_{2i}} - (y_{2i} - y_{2i-2})*\varphi\);
¡¡¡¡¡¡¡¡¡¡¡¡\EndIf
¡¡¡¡¡¡¡¡¡¡¡¡\State // Step c. Solve \({{A'}_{2i-1}}\)
¡¡¡¡¡¡¡¡¡¡¡¡\If{\(x_h > x_p\)}
¡¡¡¡¡¡¡¡¡¡¡¡¡¡¡¡\State \(c \gets (y_{2i-1}-y_{2i-2})/(x_{2i-1}-x_{2i-2})\);
¡¡¡¡¡¡¡¡¡¡¡¡\Else
¡¡¡¡¡¡¡¡¡¡¡¡¡¡¡¡\State \(c \gets (y_{2i-1}-y_{2i})/(x_{2i-1}-x_{2i})\);
¡¡¡¡¡¡¡¡¡¡¡¡\EndIf
¡¡¡¡¡¡¡¡¡¡¡¡\State \(x'_{2i-1} \gets (x_h+k*(c*x_{2i - 1} + y_h - y_{2i - 1}))/(1+c*k)\);
¡¡¡¡¡¡¡¡¡¡¡¡\State \(y'_{2i-1} \gets (y_{2i - 1}+c*(k*y_h + x_h - x_{2i - 1}))/(1+c*k)\);
¡¡¡¡¡¡¡¡¡¡¡¡\State // Step d. Check and decide the value
¡¡¡¡¡¡¡¡¡¡¡¡\If{\({x'_{2i-1}}\in ({{x}_{2i-2}},{{x}_{2i}})\)}
¡¡¡¡¡¡¡¡¡¡¡¡¡¡¡¡\State //Change the node \({{A}_{2i-1}}\)
¡¡¡¡¡¡¡¡¡¡¡¡¡¡¡¡\State \(x_{2i-1} \gets x'_{2i-1}\);
¡¡¡¡¡¡¡¡¡¡¡¡¡¡¡¡\State \(y_{2i-1} \gets y'_{2i-1}\);
¡¡¡¡¡¡¡¡¡¡¡¡\EndIf
¡¡¡¡¡¡¡¡\EndIf
¡¡¡¡\EndFor
\EndIf
\State // Do the traditional interpolation for the transformed nodes
\For{\(i=0:n-1\)}
¡¡¡¡\If{\(x\in [x_i,x_{i+1}]\)}
¡¡¡¡¡¡¡¡\State \Return \((x-x_{i+1})/(x_i-x_{i+1})*y_i+(x-x_i)/(x_{i+1}-x_i)*y_{i+1}\);
¡¡¡¡\EndIf
\EndFor
\end{algorithmic}
\end{algorithm}

\begin{algorithm}[H]
\caption{Golden curve interpolation}
\label{alg:gecurve}
\hspace*{0.02in} {\bf Input:}
\({A_0}({x_0},{y_0})\), \({A_1}({x_1},{y_1})\), \(\cdots\) , \({A_n}({x_n},{y_n})\), the derivative \({{k}_{0}}\) at the node \(A_0\), a given \(x \in [{x_i},{x_{i + 1}}]\)\\
\hspace*{0.02in} {\bf Output:}
the value of \(p(x)\)
\begin{algorithmic}[1]
\State // Do the golden extension transform
\State Create a sequential node list \(L=\{A_0\}\);
\For{\(i=0:n-1\)}
¡¡¡¡\State \(x_h \gets x_i + (x_{i + 1} - x_i)*\varphi\);
¡¡¡¡\State \(y_h \gets y_i + (y_{i + 1} - y_i)*\varphi\);
¡¡¡¡\State \(d \gets Length(L)-1\);
¡¡¡¡\State \(sum \gets 0\);
    \For{\(j=0:d-1\)}
    \State // The \(L(j).x\) represents the x-coordinate of the node whose sequence number is \(j\) in \(L\)(the sequence number of the nodes in \(L\) starts with 0)
¡¡¡¡\State \(sum \gets sum + (-1)^{d-j}*\displaystyle\frac{L(j+1).y-L(j).y}{L(j+1).x-L(j).x}\);
\EndFor
    \State \(t \gets \displaystyle\frac{1}{2}\Bigg(\displaystyle\frac{y_{i+1}-y_i}{x_{i+1}-x_i}-(-1)^{d+1}*k_0\Bigg)-sum\);
¡¡¡¡\State \(x_{i+0.5} \gets \displaystyle\frac{(y_{i+1}\!-\!y_i)\!*\!(y_h\!-\!y_i+t\!*\!x_i)+x_h\!*\!(x_{i+1}\!-\!x_i)} {t\!*\!(y_{i+1}\!-\!y_i)+x_{i+1}\!-\!x_i}\);
¡¡¡¡\If{\(x_{i+0.5}\in (x_i,x_{i+1})\)}
¡¡¡¡¡¡¡¡\State \(y_{i+0.5} \gets y_i+t*(x_{i+0.5}-x_i)\);
¡¡¡¡¡¡¡¡\State Add the node \((x_{i+0.5},y_{i+0.5})\) to the end of the list \(L\);
¡¡¡¡\EndIf
\EndFor
\State // Do the traditional interpolation for the transformed nodes
\State \(d \gets Length(L)-1\);
\For{\(i=0:d-1\)}
¡¡¡¡\If{\(L(i).x \le x \le L(i+1).x\)}
¡¡¡¡¡¡¡¡\State \Return \[\begin{split}
    &{\left(\frac{x - L(i+1).x}{L(i).x - L(i+1).x}\right)}^{2}*L(i).y + \\
    &{\left(\frac{x - L(i).x}{{L(i+1).x - L(i).x}}\right)}^{2}*L(i+1).y + \\
    &\frac{(x - L(i).x)*(x - L(i+1).x)}{L(i).x - L(i+1).x}\Bigg[2*\\
    &\sum_{j=0}^{i-1}\Bigg((-1)^{i-j+1}*\frac{L(j+1).y - L(j).y}{L(j+1).x - L(j).x}\Bigg) + \\
    &{(-1)^i}*{k_0} - \frac{2*L(i).y}{L(i).x - L(i+1).x}\Bigg];
    \end{split} \]
¡¡¡¡\EndIf
\EndFor
\end{algorithmic}
\end{algorithm}
\end{appendices}

\bibliographystyle{cag-num-names}
\bibliography{refs}

  \end{document}